\documentclass[fleqn,usenatbib]{mnras}

\usepackage{newtxtext,newtxmath}

\usepackage[T1]{fontenc}

\DeclareRobustCommand{\VAN}[3]{#2}
\let\VANthebibliography\thebibliography
\def\thebibliography{\DeclareRobustCommand{\VAN}[3]{##3}\VANthebibliography}

\usepackage{graphicx}	
\usepackage{amsmath}	
\usepackage{natbib}
\usepackage{subcaption} 
\usepackage[british]{babel}
\usepackage{placeins}
\usepackage{longtable}
\usepackage{booktabs} 
\usepackage[normalem]{ulem}

\title[Accretion vs ionised ejection in the Coronet cluster]{The Relationship between Accretion and Ionised Ejection  among \\ Young Stellar Objects in the Coronet Cluster}

\author[Arpan Ghosh et al.]{Arpan Ghosh$^{1}$\thanks{E-mail: a.ghosh@irya.unam.mx; 19aghosh91@gmail.com}, 
Roberto Galv\'an-Madrid$^{1}$, 
Johanan Ramírez-Arellano$^{1}$, 
Carlos Carrasco-González$^{1}$, 
\newauthor
Gráinne Costigan$^{2}$, 
Suzanne Ramsay$^{3}$, 
Carlo Manara$^{3}$, 
Jan Forbrich$^{4}$,
\newauthor
Hauyu Baobab Liu$^{5,6}$, 
Michihiro Takami$^{7}$ 
\\
$^{1}$Universidad Nacional Aut\'onoma de M\'exico, Instituto de Radioastronom\'ia y Astrof\'isica, 58090 Morelia, Michoac\'an, M\'exico\\
$^{2}$Grünenthal GmbH\\
$^{3}$European Southern Observatory, Karl-Schwarzschild-Strasse 2, 85748 Garching bei München, Germany\\
$^{4}$Centre for Astrophysics Research, University of Hertfordshire, College Lane, Hatfield AL10 9AB, UK\\
$^{5}$Department of Physics, National Sun Yat-Sen University, No. 70, Lien-Hai Road, Kaohsiung City 80424, Taiwan, R.O.C.\\
$^{6}$Center of Astronomy and Gravitation, National Taiwan Normal University, Taipei 116, Taiwan, R.O.C.\\
$^{7}$Institute of Astronomy and Astrophysics, Academia Sinica, 11F Astronomy-Mathematics Building, No.1, Sec. 4, Roosevelt Rd, Taipei 10617, Taiwan, R.O.C.\\
}

\date{Accepted XXX. Received YYY; in original form ZZZ}

\pubyear{\the\year{}}

\begin{document}
\label{firstpage}
\pagerange{\pageref{firstpage}--\pageref{lastpage}}
\maketitle

\begin{abstract}
We present results from a coordinated, multi-epoch near-infrared and centimeter radio survey of young stellar objects (YSOs) in the Coronet, aimed at probing the connection between 
mass accretion and ionised mass loss. Using VLT-KMOS, we detect Br$\gamma$ emission in 5 of the 26 targets, which also exhibit 3.3-cm continuum emission in VLA images, consistent with partially ionised 
jets. For seven additional sources, stringent flux upper limits were obtained. The derived accretion and ionised mass-loss rates for class~{\sc I} and class~{\sc II} YSOs follow a sublinear correlation 
$\dot{M}_{\mathrm{ion}} \propto \dot{M}_{\mathrm{acc}}^{0.3}$, consistent with previous results for class~{\sc II} YSOs but extended here to earlier stages. Multi-epoch observations reveal modest variability 
in both tracers but no clear 
temporal correlation between accretion and ejection within timescales of a few months. The ratio $\dot{M}_{\mathrm{ion}}/\dot{M}_{\mathrm{acc}}$ shows
an anti-correlation with $\dot{M}_{\mathrm{acc}}$, increasing with time from class~{\sc I} YSOs to class~{\sc II} YSOs,
suggesting an increase in jet-launching efficiency or ionisation fraction with evolution. These findings support a direct connection between accretion and outflow across the $\sim$ Myr timescale of YSO evolution, while highlighting the complexity of their short-term interplay.

\end{abstract}




\section{Introduction} \label{sec:intro}
The mass assembly during star formation occurs through the accretion of material from a circumstellar disk to the central object \citep{Hartmann2016}. During the earliest class~{\sc 0} and {\sc I} stages of Young Stellar Object evolution \citep[YSO,][]{Adams1987, Andre1993}, the protostars and their disks are also surrounded by dense envelopes, which are cleared out within a Myr \citep{Dunham14_PPVI}.
The later YSO stages (class~{\sc II} and {\sc III}) are marked by the evolution of their disks, with a gradual decline in their gas content and accretion
rates over a few Myr \citep{Armitage2003, WillamsCieza2011, Manara23_PPVII,Gaidos2025}.

A fraction of the accreted material is ejected in the form of jets and disk-winds, which are crucial for the release of specific angular momentum, as well as for disk dispersal \citep{Frank14_PPVI,Pascucci2023}. 
In standard theoretical models, the ejection and accretion of material are coupled through magnetohydrodynamical (MHD) processes \citep{Shu1994,Pudritz05_PPV}. Therefore, a common expectation is that accretion and ejection will be coupled in time. However, detailed theoretical modelling of these processes across YSO evolution is challenging \citep[e.g.,][]{Romanova18,Zhu25}.  
Furthermore, YSOs show evidence of time variability within timescales much shorter than their evolutionary timescale 
in their accretion rates, ejected jets, and circumstellar material \citep[e.g.,][]{MoralesCalderon2011,Fischer23_PPVII,Lora24}.  

A joint characterisation of accretion and ejection processes across YSO evolutionary types requires multiwavelength data. 
Emission of hydrogen (H{\sc i}) recombination lines such as Br$\gamma$, emitted in the accretion shock around the stellar surface, have become the standard to measure accretion  \citep{FolhaEmerson2001,Muzerolle2001}. 
In the least embedded YSOs, jets and disk-winds are commonly studied via optical lines, e.g., [O{\sc i}]  \citep{Hartigan1995,Fang2018,Nisini2018, Banzatti2019}, as well as with near-infrared molecular hydrogen H$_2$ emission  \citep{2006ApJ...641..357T,2017MNRAS.465.3039C,2020MNRAS.492..294G}. Radio continuum emission at centimeter wavelengths is known to be an effective tracer of the partially ionised base of jets \citep{Rodriguez2014,2018A&ARv..26....3A}. This tracer is particularly valuable in the more embedded class~{\sc 0/I} YSOs \citep{Tychoniec2018}. However, other physical processes can contribute to the observed centimetre emission. Non-thermal (gyro)synchrotron emission from an active stellar magnetosphere is expected to dominate in class~{\sc IIIs} and could contaminate the emission of less evolved YSOs \citep[e.g.,][]{2014ApJ...780..155L,Forbrich21}.

In this paper, we report the results of a coordinated monitoring of YSOs in the R Corona Australis (R CrA) star forming region, using Br$\gamma$ and centimeter continuum as tracers of accretion and ejection, respectively. 
Recent studies by \citet{Rota2025,Garufi2025,Rota2024} have explored the relationship between accretion and ejection using these tracers, but they have focused on class~{\sc II} YSOs, and their observations were not quasi-simultaneous. The {\it Coronet} cluster in R CrA is an excellent target because it is nearby (152 pc; \citealt{Galli2020}, see also \citealt{Dzib2018}) and remains
embedded in its natal cloud \citep{2011ApJ...736..137S}, hosting a large concentration of YSOs at different evolutionary stages \citep{2005A&A...429..543N,Aurora2008, 2009ApJ...700.1609M,2011ApJS..194...43P,Cazzoletti2019}. The Coronet has also been studied in the radio continuum \citep{2008A&A...486..799M,Choi2008,2014ApJ...780..155L}, including monitoring campaigns with X-ray observations \citep{Forbrich2007}. 

The paper is arranged as follows. Section \ref{sec:obs} describes the observations and data analysis, followed by the results in Section \ref{sec:results}. The implications of our results are discussed in Section \ref{sec:discussion}, and Section \ref{sec:conclusions} presents our conclusions.

\section{Observations and Data Analysis}\label{sec:obs}

\subsection{Near-infrared data}

We have selected a sample of 26 YSOs within the Coronet cluster, spanning  the evolutionary range from class {\sc I} to class {\sc III}. The YSOs were selected from those reported by 
\citet{2011ApJS..194...43P} in their study using \textit{Spitzer}. The selected YSOs were observed with the  K-band Multi Object Spectrograph \citep[KMOS,][]{Sharples2013} 
on the Very Large Telescope (VLT), using the K-band setup, which provides a wavelength coverage between 1.934 and 2.460 $\mu$m
at a resolution of $R \sim$ 4200. The observations were carried 
out between April 27 and July 22, 2014, as 
part of proposal ID 093.C-0657 (P.I. Galván-Madrid). The target-of-opportunity (ToO) mode was used, triggered by observations taken with the Very 
Large Array. The data were reduced using the KMOS version 2.7.3 pipeline within ESOReflex \citep{2013A&A...558A..56D, 2013A&A...559A..96F}. 
The 1D spectrum was extracted by averaging over a $10 \times 10$ pixel region of the spatial axes centred on the source peak, chosen to maximize the S/N 
and ensure uniform aperture losses across epochs with varying seeing conditions. For epochs where the source was not well centred within the IFU field of 
view, a smaller region was adopted.
The spatially integrated spectra from the output KMOS datacubes were calibrated using photometry obtained from the 2MASS \citep{2MASS2006} and VISIONS \citep{VISIONS} surveys. 
In a way similar to \citet{2021A&A...650A..43F}, this latter procedure was adopted to improve the spectro-photometry of the original KMOS data.

\subsection{Radio Data} \label{sec:radio-data}

Observations were conducted with the NRAO's Karl G. Jansky Very Large Array (VLA). Partial results of the monitoring program, which ran from 2012 to 2015, have been published in \citet{2014ApJ...780..155L} and \citet{GM2014}. The full results will be presented in Ramírez-Arellano et al. (in prep.)
During the months of the KMOS program, the VLA observed for ten epochs: seven in the X band (3.3 cm) and three in the Ku band (2.1  cm). In this paper, we make use of the higher quality X-band data from those epochs, as well as the image created by concatenating the visibilities of 32 X-band epochs from the entire program.

The VLA observations were carried out in single-pointing mode, centered at 
$\alpha(\mathrm{J2000}) = 19^h ~01^m ~48.0^s$,  
$\delta(\mathrm{J2000})=-36^\circ ~57' ~59.00''$. 
Data were calibrated using the Common Astronomy
Software Applications package, version 6.5.4 \citep{CASA2022}. 
Standard calibration and imaging steps were followed \citep[see][]{2014ApJ...780..155L}. The phases of the visibilities in each epoch were self-calibrated, and individual-epoch images were created with these. Also, 
deep images were made from the concatenated visibilities of all epochs. All the  images used in this paper were produced using the \texttt{tclean} task within \texttt{CASA}, utilizing $Briggs$ weighting with $\mathrm{robust}=0$. The central frequency of the images is 9.0 GHz (3.3 cm).

The most sensitive image is the one created with the concatenated visibilities. 
The rms noise prior to the correction of the primary-beam response is $9\ \mu \mathrm{Jy\ beam^{-1}}$. The FWHM beamsize of this image is $\theta_{maj}\times\theta_{min}=1''.85\times0''.78$, with a position angle (P.A.) = $4.2^\circ$.

\begin{table}
\centering
\caption{Summary of the parameters of the Br$\gamma$ detected sources. Sources with upper limits in both Br$\gamma$ and 3.3 cm are labelled `Ul'.}
\label{tab:source_data}
\setlength{\tabcolsep}{2pt}
\begin{tabular}{l c c c c c c c c}
\hline
Source & RA & DEC & YSO & Br$\gamma$ & 3.3cm & Mass & Radius & $A_V$ \\
       & (J2000) & (J2000) & Class & & Radio & ($M_\odot$) & ($R_\odot$) & (mag) \\
\hline
IRS1   & 19:01:50.705 & -36:58:09.55 & \textsc{I}   & Em  & Yes & 0.40 & 6.00 & 30.0 \\
IRS2   & 19:01:41.582 & -36:58:31.07 & \textsc{I}   & Em  & Yes & 1.40 & 2.90 & 20.0 \\
CrA43  & 19:01:58.564 & -36:57:08.23 & \textsc{I}   & Em  & Yes & 3.54 & 2.32 & 16.5 \\
IRS6   & 19:01:50.480 & -36:56:37.90 & \textsc{II}  & Abs & Yes & 0.20 & 2.32 & 29.0 \\
TCrA   & 19:01:58.784 & -36:57:49.78 & \textsc{II}  & Em  & Yes & 2.25 & 1.14 & 2.5 \\
CrA16  & 19:01:33.877 & -36:57:44.71 & \textsc{II}  & Em  & Yes & 0.32 & 1.22 & 16.6 \\
JVLA1  & 19:01:34.917 & -37:00:56.84 & \textsc{III} & Abs & Yes & 1.34 & 1.94 & 1.4 \\
CrA26  & 19:02:06.833 & -36:58:41.43 & \textsc{II}  & Ul  & Ul  & 0.18 & 2.10 & 1.2  \\
Peterson1 & 19:01:32.345 & -36:58:02.97 & \textsc{II} & Ul & Ul & 0.28 & 1.21 & 14.7  \\  
Peterson6 & 19:01:53.775 & -37:00:33.80 & \textsc{II} & Ul & Ul & 0.10 & 0.47 & 0.2 \\
IRS10     & 19:02:04.109 & -36:57:00.72 & \textsc{II} & Ul & Ul & 0.16 & 1.68 & 9.76 \\   
\hline
\end{tabular}
\begin{flushleft}
\textit{Notes.} Stellar parameters are from \citet{Nisini2005} and \citet{Dong2018}. For CrA43, Peterson1, Peterson6, IRS10 and CrA26, parameters were obtained from SED fitting using \texttt{SEDFITTER} \citep{Robitaille}.
\end{flushleft}
\end{table}

\subsection{Flux measurements}

We inspected  the flux-calibrated, spatially averaged KMOS spectra of each epoch, looking for the Br$\gamma$ line at 2.166 $\mu$m. 
We detected Br$\gamma$ in seven out of the 26 targeted YSOs. Among them, the line was found to be in emission in five YSOs and in absorption in 
the 
remaining two (see Table \ref{tab:source_data}). To measure the Br$\gamma$ fluxes, we fitted a model that includes a Gaussian for the 
Br$\gamma$ line and a first-order polynomial for the adjacent continuum. For this, we used the  \texttt{LevMarLSQFitter} within the \texttt{astropy modeling} 
package \citep{astropy2022}.  
The line flux was measured as the area under the fitted Gaussian, along with the corresponding error propagation. This procedure was performed for the individual 
epochs and for the median-combined spectrum of each source. 
Of the remaining 19 targets, two (JVLA4 and Peterson4) had no detectable continuum within the IFU field of view, one (CrA19) had a coordinate mismatch with the radio data, and two with extended radio emission 
(IRS7E and IRS7W) were excluded. The remaining 14 sources show no Br$\gamma$ detection. These comprise seven class~{\sc III}, six class~{\sc II}, and one class~{\sc I} YSOs. The 1D spectra of the seven 
class~{\sc II} and class~{\sc I} sources from individual epochs 
were averaged and combined to produce a single spectrum. The upper limit on their Br$\gamma$ flux was then estimated following the procedure of \citet{2021A&A...650A..43F}, modified as follows:
\begin{equation}
F_{\mathrm{upper}} = 3 \times \sigma_{\mathrm{rms}} \times \sqrt{N_{\mathrm{chan}}} \times \Delta v,
\end{equation}
where $N_{\mathrm{chan}}$ corresponds to the number of channels across the line width, given by $\lambda_{\mathrm{line}}/R$, with $\lambda_{\mathrm{line}}$ as the central Br$\gamma$ wavelength, $R$ the KMOS $K$-band resolution, and $\Delta v$ as the channel spacing.

VLA data obtained nearly simultaneously\footnote{Since the VLT and VLA epochs were not observed at the exact same time, we consider observations separated by up to ten days as 
near-simultaneous.} with the KMOS observations were used to investigate temporal correlations between the two. 
Photometry in the radio images was obtained using the \texttt{imfit} task within \texttt{CASA}, which performs Gaussian fitting of the source emission, including error estimation. We take the fitted peak intensity as the source flux. This ensures that the radio flux corresponds to the same arcsecond scale as the KMOS IFU measurement. 
Also, time-averaged photometry was obtained from the \texttt{imfit} results in the concatenated image and from averaging the results in the individual epochs matching the KMOS program (see Section \ref{sec:radio-data}).

\begin{figure}
\centering
\includegraphics[width=0.48\textwidth]{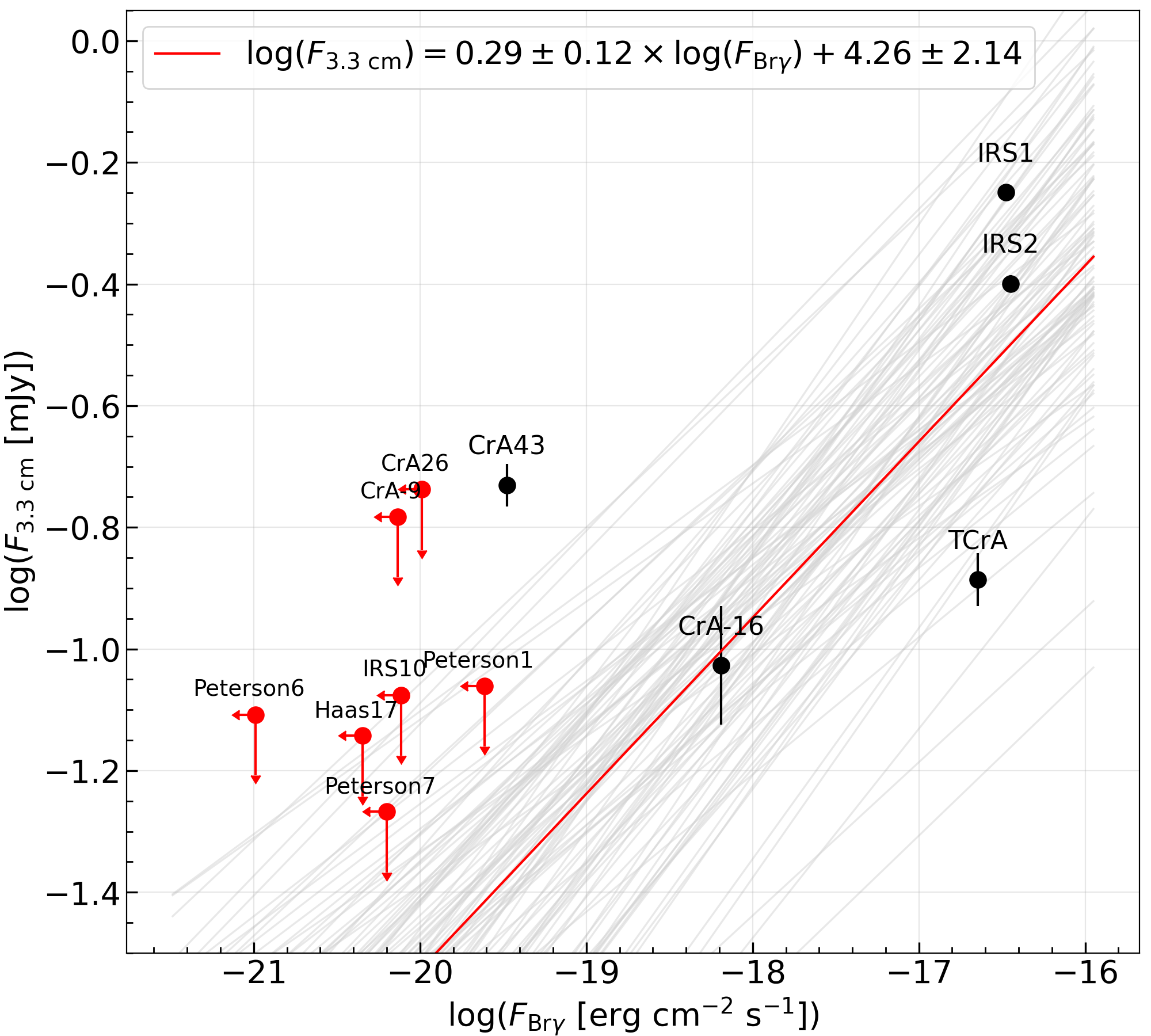}
\caption{{\small Relationship between the Br$\gamma$ and the 3.3 cm radio continuum fluxes for the class {\sc I} and {\sc II} sources. Detected sources are denoted by black symbols and upper limits by red symbols. Grey curves illustrate posterior regression lines consistent with the $68\%$ credible interval of the slope; the median relation is shown in red.
}} 
\label{fig:fig1}
\end{figure}

\section{Results} \label{sec:results}

\subsection{Relationship between Br\texorpdfstring{$\gamma$}{gamma} and 3.3 cm radio fluxes} \label{secc:fluxvsflux}

Table \ref{tab:source_data} lists the Br$\gamma$ and radio detections of the YSOs. In the per-epoch images, the radio continuum emission from 
CrA16 was not detected. However,in the concatenated radio image, this source has a detection at $\approx 94\pm21~\mu$ Jy beam$^{-1}$. The rest of the Br$\gamma$ detected radio sources are detected both in the concatenated image and in individual epochs. 
Among the YSOs, only JVLA1 and IRS6 have Br$\gamma$ in absorption, whereas in the rest of the detections, the line is in emission. IRS6 is a binary system, and the component designated as IRS6a is the only one with detectable spectral features \citep{2005A&A...429..543N}.
In this paper, references to IRS6 specifically denote the IRS6a component. The Br$\gamma$ absorption line profile of IRS6 is not very wide; hence, we refrain from inferring 
further about the observed profile. 
JVLA1 is a class {\sc III} YSO with a K1 {\sc IV} subgiant spectral type \citep{2007A&A...475..959F}. The KMOS spectrum of JVLA1 has prominent 
absorption features in Br$\gamma$ and the CO (2--0) and (3--1) bandheads. \citet{2011ApJ...736..137S} reported minimal infrared excess in JVLA1. Therefore, the absorption lines of JVLA1 likely originate from a cool photosphere. 


\begin{figure}
    \centering
    \includegraphics[width=0.73\columnwidth]{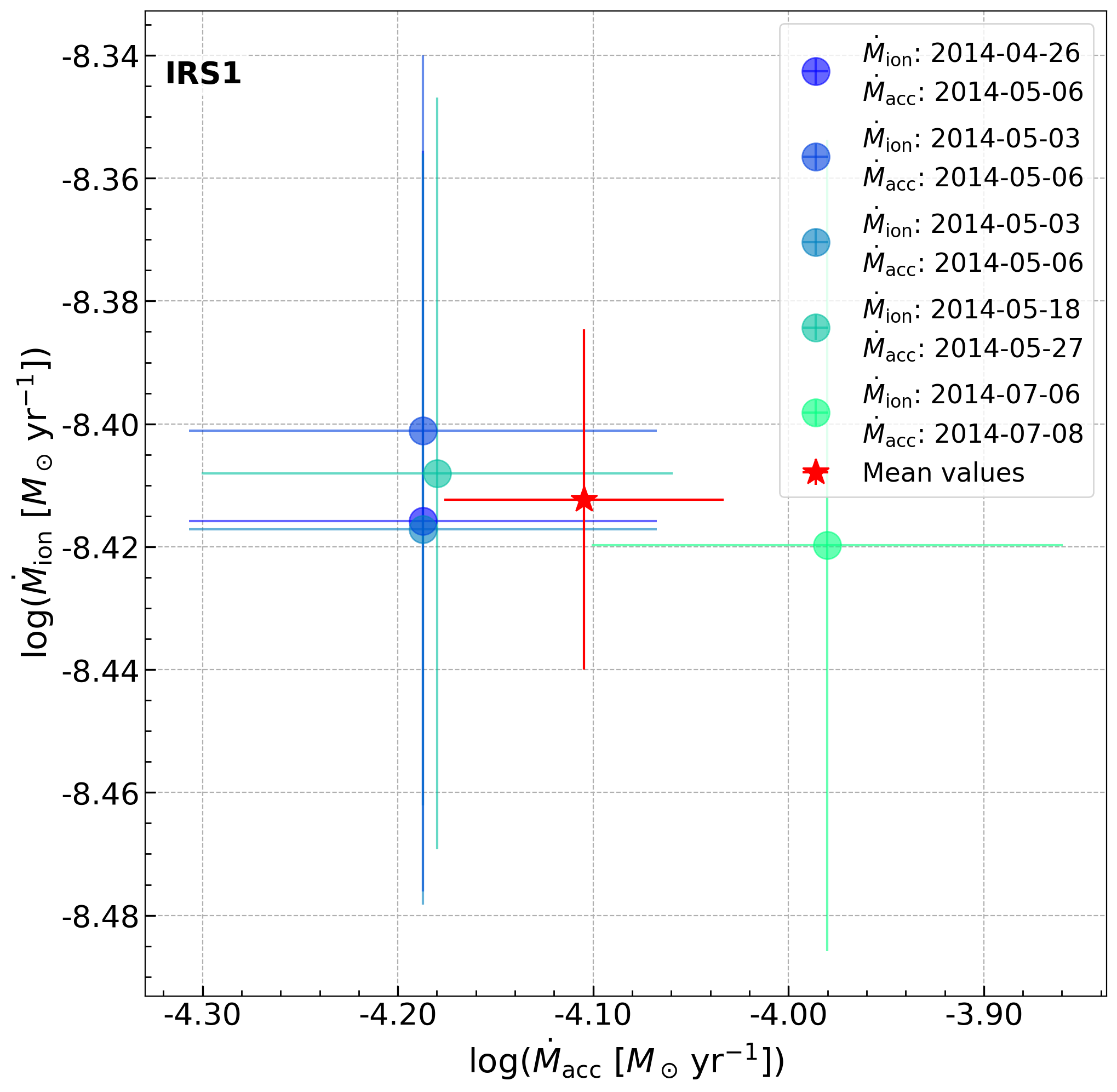}\par\vspace{-0.3cm}
    \includegraphics[width=0.73\columnwidth]{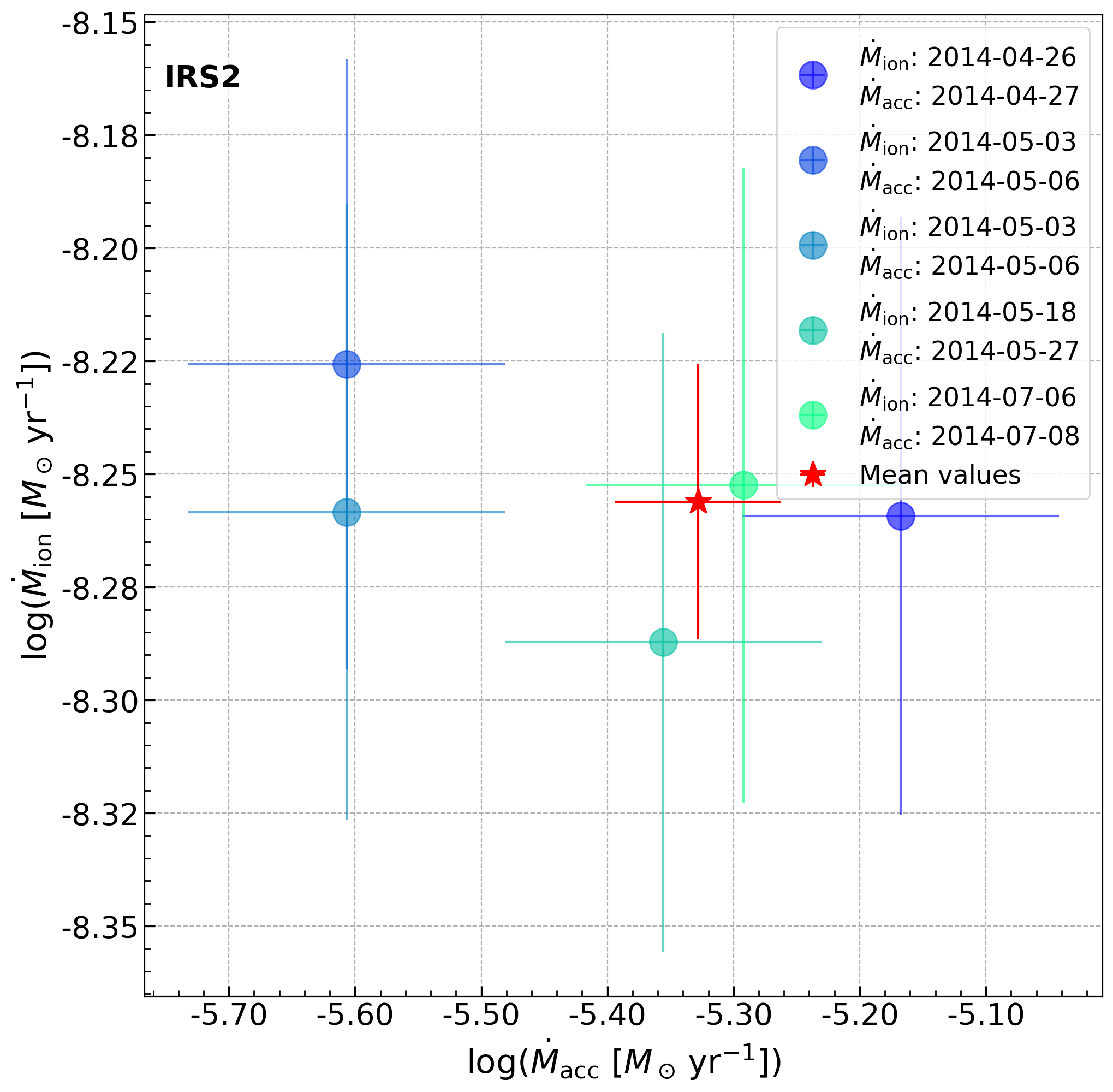}\par\vspace{-0.3cm}
    \includegraphics[width=0.73\columnwidth]{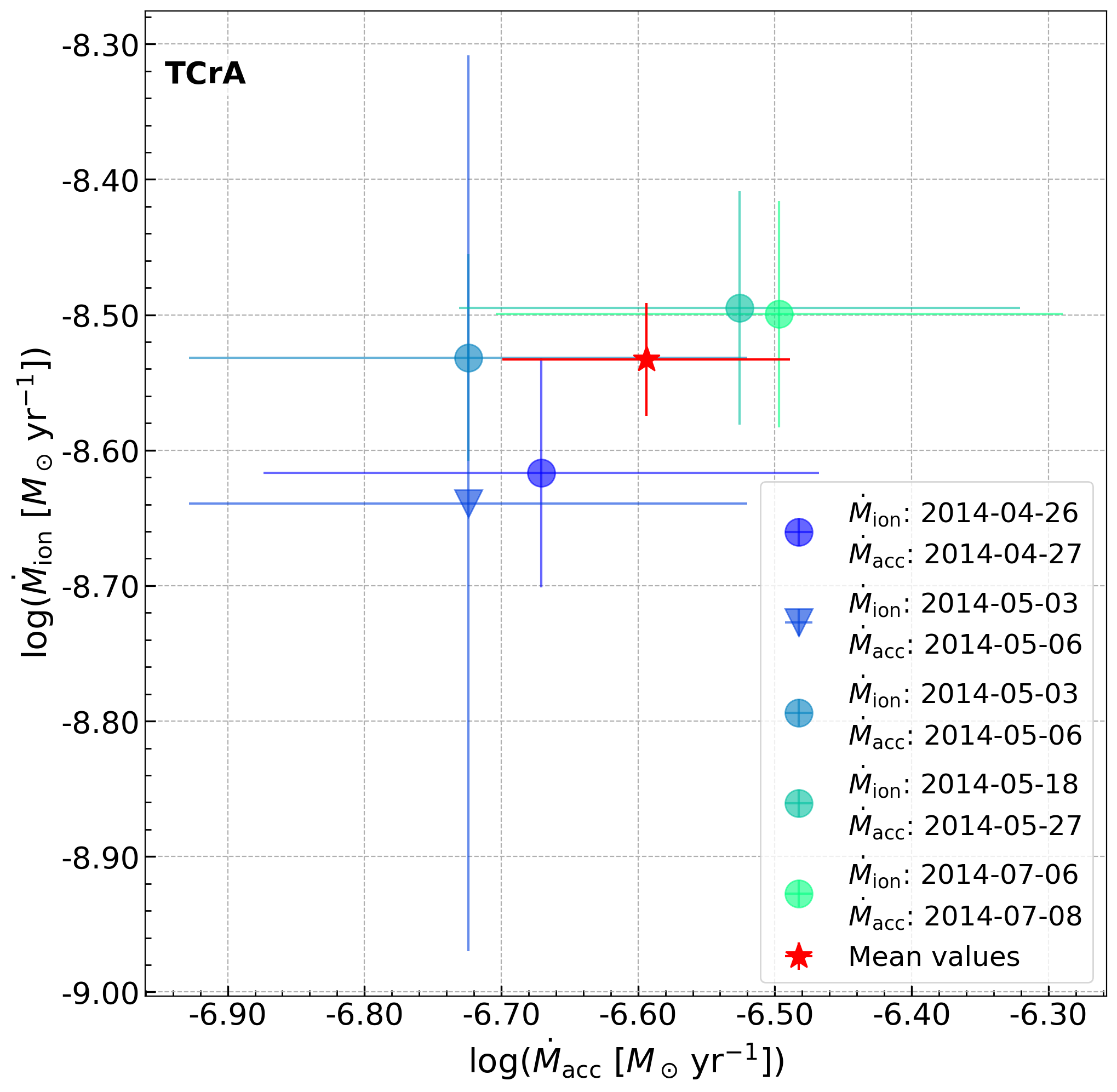}
    \caption{Temporal evolution of individual measurements of $\dot{M}_{\mathrm{acc}}$ and $\dot{M}_\mathrm{ion}$ for the two class \textsc{I} (IRS1, IRS2, top and middle panels) and one class \textsc{II} YSO 
    (TCrA, bottom) with Br$\gamma$ in emission and VLA detection in individual epochs.  
    CrA43 was excluded due to its unphysically large ejection rate compared to its accretion rate (see Section \ref{outlier}).  
    The blue and green points refer to values around the individual epochs marked in the plot legends, whereas the red star is the mean value.}
    \label{fig:fig2}
\end{figure}

The time-averaged Br$\gamma$ and 3.3 cm fluxes in our sample are shown in Figure \ref{fig:fig1}. 
We have used the Br$\gamma$ fluxes averaged over all the observation epochs and the radio fluxes from the concatenated image at 3.3 cm.  
We have employed a Markov Chain Monte Carlo (MCMC) method for the Bayesian linear regression since it is well suited for datasets containing measurements with uncertainties in both variables. 
A linear fit to the data gives the following shallow, positive relation:\footnote{Throughout the rest of this paper, radio continuum fluxes are in units of millijansky (mJy), and Br$\gamma$ fluxes are in units of erg s$^{-1}$ cm$^{-2}$.}
$\log(F_{\mathrm{3.3\,cm}}) = (4.3\pm2.1) + (0.29\pm0.12) \times \log(F_{Br{\gamma}})$.
As we will see in the following, a correlation between these two tracers is not expected for all YSO types.

In the standard scenario of disk evolution, accretion rates are expected to decrease significantly from the class~{\sc I} to the class~{\sc II} YSO stages \citep{Hartmann2016}. Consequently, if the centimeter free-free emission from YSOs is caused mainly by partially-ionised jets \citep{2018A&ARv..26....3A}, and if the strength of those jets is linked to that of accretion, a relationship between Br$\gamma$ and radio continuum is expected. This has been recently reported for class~{\sc II} YSOs \citep{Rota2024,Garufi2025}, including transition disks \citep{Rota2025}. 
However, the faint centimeter continuum from class~{\sc II} YSOs could also be due to disk photo-evaporation by EUV photons \citep{Pascucci2014,GM2014,Macias2016}, which could nevertheless be linked to accretion in a complex way via disk winds \citep{Pascucci2023}.  
In the class~{\sc III} stage, where the magnetosphere of the un-embedded young star becomes exposed, centimeter radio emission can become bright again, highly variable, and non-thermal in nature \citep{2014ApJ...780..155L,Forbrich21}. 
This non-thermal radio emission 
is not expected to be correlated with accretion phenomena. 
It is worth mentioning that class~{\sc 0/I} YSOs also exhibit non-thermal radio emission; however, only a few have been observed, e.g., IRS 5 in the Coronet cluster 
\citep{Feigelson99ARAA,2013Forbrich,2014ApJ...780..155L}. Such emission is typically identified through circular polarisation, however, it is expected that in the majority of younger YSOs non-thermal emission is  easily veiled by the free-free component along the line of sight \citep{1987Andre,2013Forbrich}.

\begin{figure*}
\centering
\includegraphics[width=0.47\textwidth]{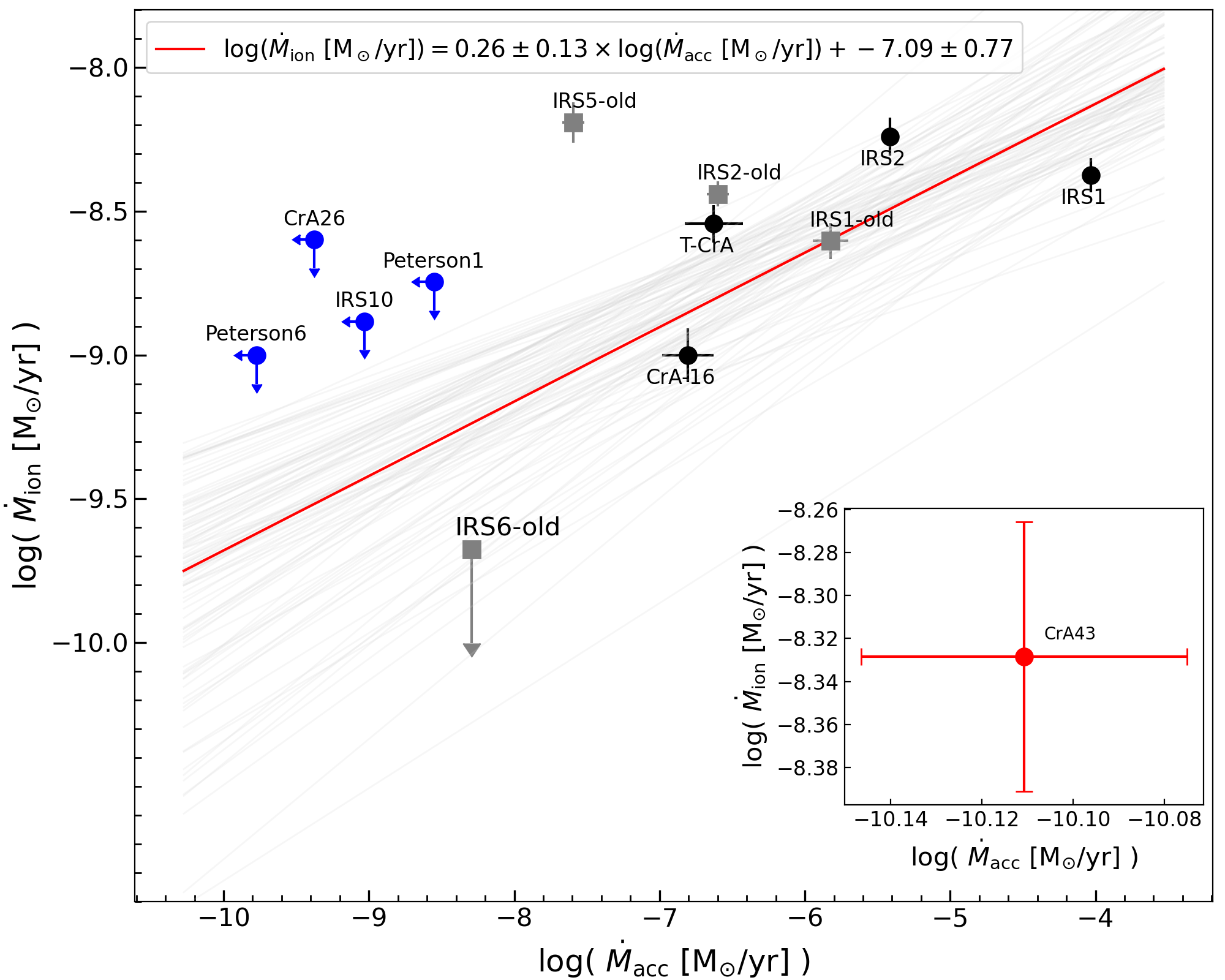}
\includegraphics[width=0.47\textwidth]{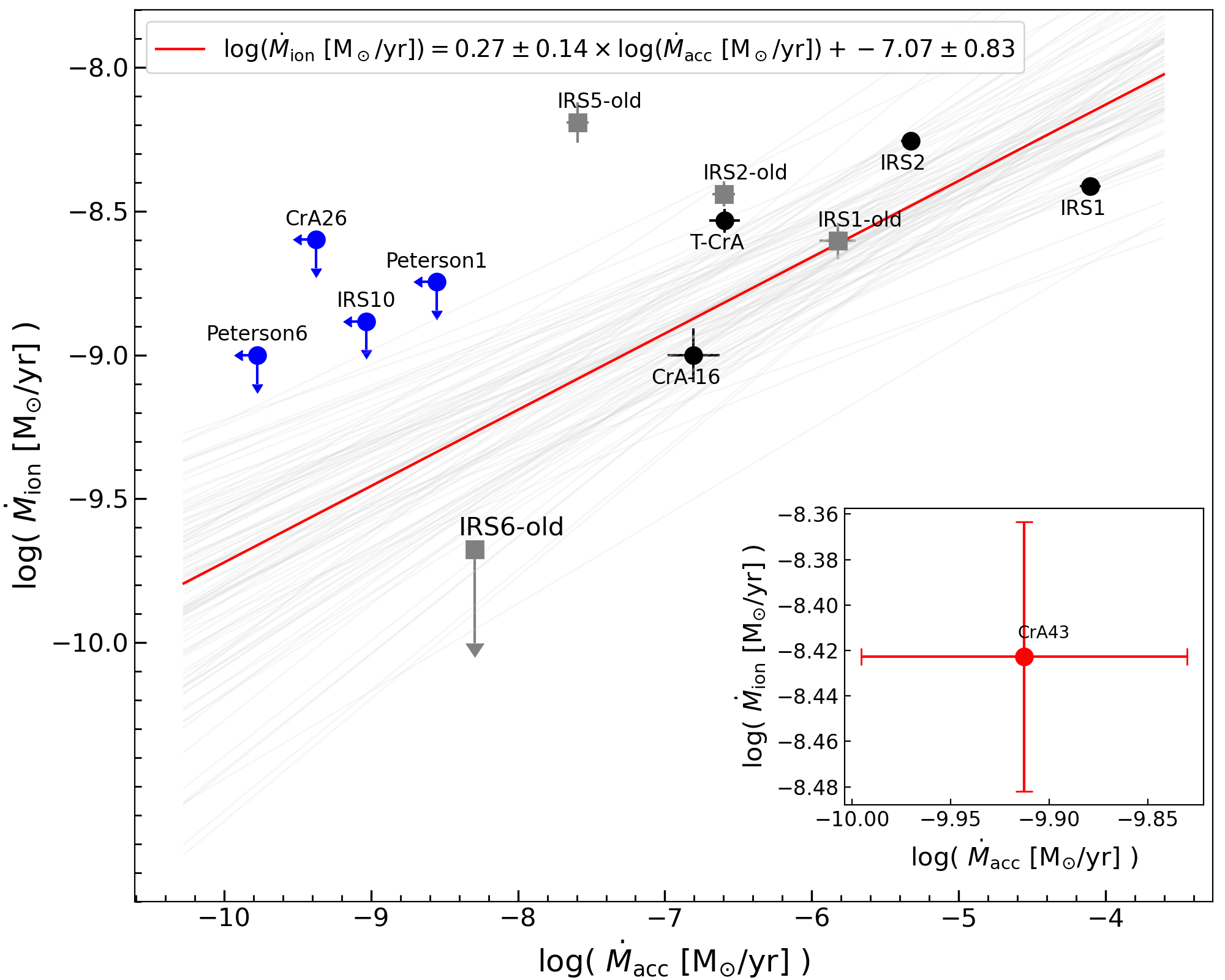}
\caption{{\small Correlation between the logarithm of the mass accretion rate and the logarithm of the ionised mass-loss rate, 
including the data from IRS1, IRS2, CrA16, TCrA, Peterson 1, Peterson 6, and CrA 26, but excluding CrA43 (inset) from the fit.  
We present two fitting scenarios: in the left panel the fit is obtained utilizing the averaged Br$\gamma$ spectra and epoch-concatenated VLA image; in the right panel the fit is obtained from averaging the near-simultaneous measurements shown in Fig. \ref{fig:fig2} for the detections, and upper limits as described in the text. The light grey lines show regression relations drawn from posterior samples restricted to the central $68\%$ credible interval of the slope; the red line indicates the posterior median.
The overplotted gray squares, marked with the suffix ``-old'', 
show previous measurements obtained from  \citet{2014ApJ...780..155L} and 
\citet{2005A&A...429..543N}. 
}}
\label{fig:fig3}
\end{figure*}

\subsection{Relationship between accretion and ionised mass-loss rates}

Br$\gamma$ fluxes were converted to the corresponding mass accretion rates ($\dot{M}_\mathrm{acc}$) using the relations of \citet{Alcala2017}. Although these relations were calibrated for class~{\sc II} YSOs, we also apply them to class~{\sc I} YSOs,
assuming that the underlying accretion engine in class~{\sc I} protostars is similar to that of more evolved class~{\sc II} systems, as supported by recent studies \citep[e.g.][]{Testi2025}.
The stellar parameters of mass 
($M_\ast$), radius ($R_\ast$), and extinction ($A_V$) are summarised in Table \ref{tab:source_data}. The conversion was carried out for five class~{\sc I} and {\sc II} 
sources with Br$\gamma$ detections and for three of the seven upper-limit YSOs with available photometry from VISIONS and {\it Spitzer}/WISE.
The Br$\gamma$ fluxes of the YSOs have been extinction corrected with corresponding $A_V$ values using the relation given in \citet{Alcala2017} and uncertainties in the fluxes, stellar masses, and radii were propagated into the accretion-rate estimates. 
Since class~{\sc I} YSOs have significant veiling\footnote{Veiling refers to the excess continuum emission 
arising from the circumstellar disk due to the accretion process \citep[e.g.,][]{CalvetGullbring98}.} in their spectra, we have followed the recipe described by \citet{2021A&A...650A..43F} to correct for this veiling in our three class~{\sc I} sources; accretion rates were computed over the allowed range of veiling values and combined to derive weighted mean $\dot{M}_\mathrm{acc}$ estimates. 

The 3.3 cm radio continuum fluxes are transformed to ionised mass-loss rates ($\dot{M}_\mathrm{ion}$) for the  eight class~{\sc I} and {\sc II} YSOs with an accretion
rate measurement or upper limit (see Table~1 in the online supplementary material). 
Five of them are VLA detections (IRS1, IRS2, CrA43, TCrA, CrA16). For CrA26, Peterson 1, and 
Peterson 6 we use $3\sigma$ from the local noise in the VLA image.    
The class~{\sc III} YSO JVLA1 has been excluded because the 
jet interpretation is not appropriate for the radio continuum of the more evolved YSOs (see Section \ref{secc:fluxvsflux}). 
We follow equation (11) of \citet{2018A&ARv..26....3A}, assuming a conical jet geometry. Disk inclination angles ($i$) for IRS1, IRS2, TCrA, CrA16, and CrA43 are obtained from \citet{Cazzoletti2019, Hsieh24}. For sources lacking inclination information, we assume $i=60^\circ$, corresponding to the median inclination angle expected for a randomly oriented sample. 
Following \citet{Rota2024}, we assume the jet to be perpendicular to the disk plane and therefore aligned with the inclination of the outer disk. The jet velocity ($v_\mathrm{jet}$) is estimated from the stellar mass $M_\ast$ using Equation~(12) of \citet{2018A&ARv..26....3A}. We conduct this analysis for both the epoch-averaged data and the individual epoch detections. The errors in the mass rates were estimated using the \texttt{uncertainties}\footnote{https://pypi.org/project/uncertainties/} Python package, propagating from the  
errors in the measured fluxes, disk inclination, jet angular size, jet velocity and electron temperature.

Figure \ref{fig:fig2} presents the temporal evolution of $\dot{M}_\mathrm{acc}$ and $\dot{M}_\mathrm{ion}$ for the three YSOs with individual-epoch measurements. CrA43, although detected in individual epochs, is excluded from further analysis because its derived $\dot{M}_\mathrm{ion}$ is unphysically larger than $\dot{M}_{\mathrm{acc}}$ (see Section~\ref{outlier}). 
The three analysed objects, IRS1, IRS2, and TCrA, 
exhibit variability of 0.12, 0.27, and 0.17 dex, respectively, from their mean rates.
No apparent correlation in time is seen between these two quantities, suggesting that a larger temporal coverage is needed \citep[ e.g.,][]{Takami2020}.  

For the time-averaged values, we performed Bayesian linear regression fits to the logarithm of $\dot{M}_\mathrm{ion}$ as a function of the logarithm of $\dot{M_\mathrm{acc}}$ under two conditions: $i)$ taking the Br$\gamma$ fluxes from the fits to the averaged spectra and the 3.3 cm fluxes from the epoch-concatenated VLA image (see Fig~1 and Fig~2 of the online supplementary material) 
and $ii)$ from averaging the $\dot{M_\mathrm{acc}}$ and $\dot{M}_\mathrm{ion}$ values of the near-simultaneous epochs, as shown in Fig. \ref{fig:fig2}, but taking the 3.3 cm flux of CrA16 from the concatenated image and the upper limit fluxes of Peterson 1, Peterson 6, and CrA26 from the 
concatenated centimeter image and Br$\gamma$ spectra. Based on the expected physical connection
between $\dot{M}_\mathrm{ion}$ and $\dot{M_\mathrm{acc}}$, we assume flat bounded priors on the slope (0-2) and intercept (-10 to 10).
Figure \ref{fig:fig3} shows the 
$\dot{M_\mathrm{acc}}$ and $\dot{M}_\mathrm{ion}$ values and fits for these two cases.  
We also converted the 3.3 cm fluxes at lower angular resolution reported in \citet{2014ApJ...780..155L} into the corresponding  $\dot{M}_\mathrm{ion}$, and overplotted them in Figure \ref{fig:fig3}, using accretion rates from \citet{2005A&A...429..543N}. 

The obtained fits to our data are 
$\log(\dot{M}_\mathrm{ion}) = 
(-7.1\pm0.8) + (0.26\pm0.13) \times \log(\dot{M_\mathrm{acc}})$ 
for case $i$, and 
$\log(\dot{M}_\mathrm{ion}) = 
(-7.1\pm0.8) + (0.27\pm0.14) \times \log(\dot{M_\mathrm{acc}})$ 
for case $ii$.
This analysis aimed to assess the impact of temporal variability in our coordinated observations, as YSOs can display short timescale behaviour (e.g. dipper or burster variability) capable of significantly affecting the inferred relationships. The observed variations in $\dot{M}_{\rm ion}$ and $\dot{M}_{\rm acc}$ across the monitoring epochs were modest of the order of $\sim 0.1$ dex, and do not affect the resulting slope of the relation.

\section{Discussion} \label{sec:discussion}

\subsection{Ionised ejection with respect to accretion}

\begin{figure}
\centering
\includegraphics[width=0.48\textwidth]{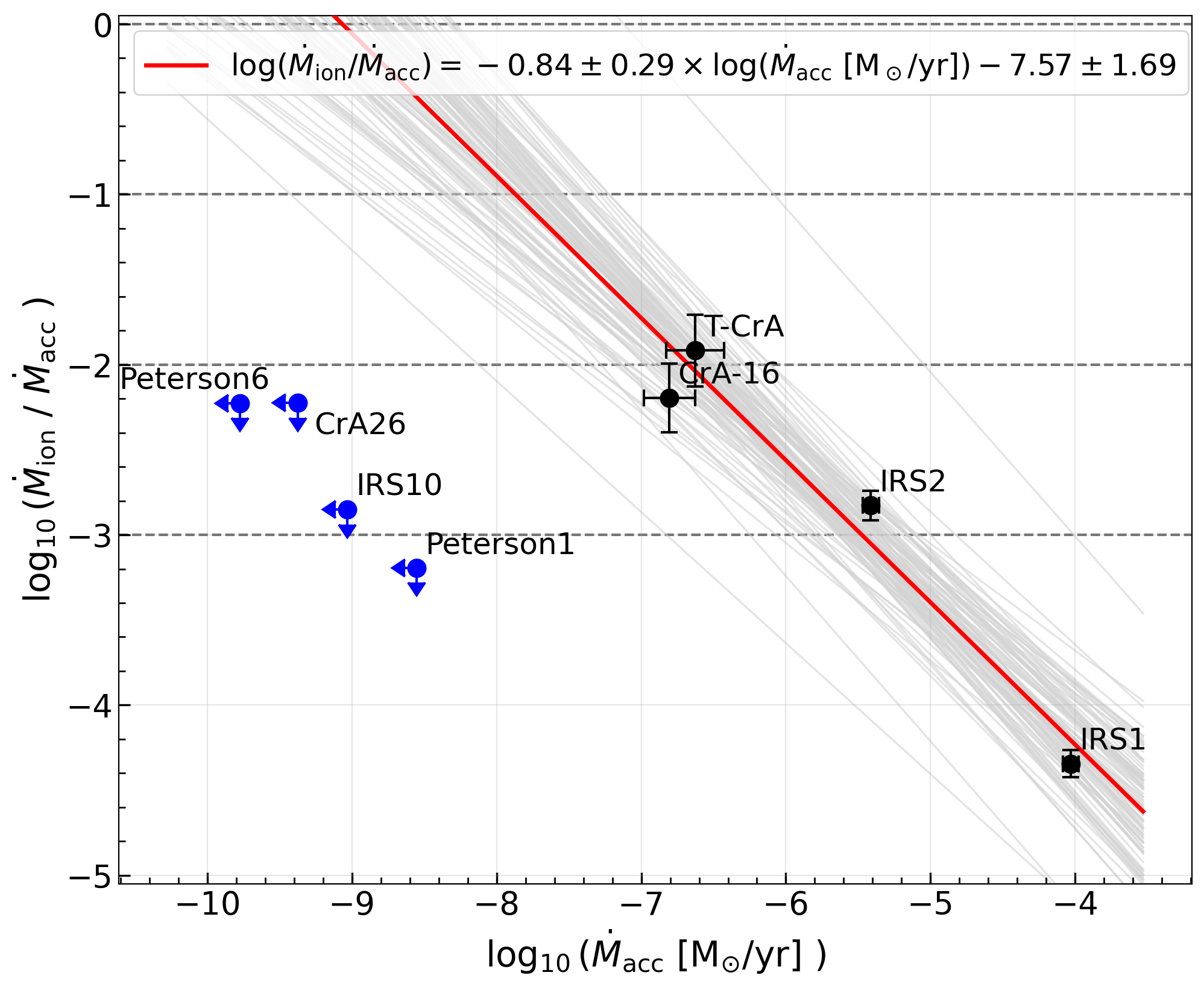}
\caption{{\small Logarithm of the ratio of the ionised mass-loss rate to the mass-accretion rate, as a  
function of the latter. The horizontal lines mark efficiency factors $\xi_\mathrm{ion} =$ 0.001, 0.01, 0.1, and 1. The red and grey lines are defined as in Fig. \ref{fig:fig2}. }}
\label{fig:fig4}
\end{figure}

Figure~\ref{fig:fig4} shows an anti-correlation between the logarithmic ratio $\dot{M}_\mathrm{ion}/\dot{M}_\mathrm{acc}$ as a function of the logarithm of  $\dot{M}_\mathrm{acc}$.   
A linear fit gives 
$ \log ( \dot{M}_\mathrm{ion}/ \dot{M_\mathrm{acc}}) = (-7.6\pm1.7) - (0.84\pm0.29) \times \log(\dot{M_\mathrm{acc}})$. 

Considering that $\dot{M}_\mathrm{ejection}/ \dot{M_\mathrm{acc}}$ probes the efficiency $\xi$ of the ejection process 
compared to that accreted by the protostar, in our case, the free-free emission is only sensitive to the ionised component of the ejected material, $\xi_\mathrm{ion}$, not to the neutral atomic \citep{Nisini2018} or molecular \citep{TomRay2023} components. 
The $\xi_\mathrm{ion}$ values in our sample are consistent with those reported by \citet{Rota2024, Rota2025}, who also used the free-free continuum for a sample of class~{\sc II} YSOs mostly located in the Taurus star formation region. Our $\xi_\mathrm{ion}$ values are also consistent with those of \citet{Nisini2018}. However, their ejection rates were derived from the high velocity component of the [O {\sc i}] line in a sample of 131 class~{\sc II} YSOs in several star forming regions. 
\citet{Rota2025} showed that, although full discs (FDs) typically accrete at higher rates than transition discs (TDs), the accretion-wind scaling is shallower in TDs, implying a reduced apparent efficiency. 
The four sources for which we could perform the full analysis are two class IIs and, for the first time, two class I YSOs. 

Several reports of rings and gaps in class~{\sc I} disks have been presented  \citep[e.g.,][]{Sheehan2020,Hsieh24}, but the details of mass transport in the disks of embedded YSOs are far from understood. 
In class~{\sc II} YSOs, studies on magneto-centrifugal launching models have shown that $\xi$ is inversely proportional to the magnetic lever arm $\lambda^{-1}$ \citep{Pelletier1992,Hartmann2016,Pascucci2023}. This implies that lower $\xi$ values correspond to larger $\lambda$, thus resulting in more compact jet launching regions. The lower $\xi_{ion}$ values 
for class~{\sc I} YSOs in our sample may therefore point to narrower launching zones or different magnetic field morphologies, a scenario pointed out by \citet{Nisini2018}.      

On the other hand, 
protostellar jets are known to have a nested structure, consisting of high-velocity ionised winds enclosed by layers of atomic and molecular jets with decreasing velocity gradients \citep[e.g.,][]{Nisini2024,Bacciotti2025}. 
Therefore, another possible interpretation of the trend shown in Fig. \ref{fig:fig4} considers the chemical and ionization state of the ejected material, which can be either atomic ionised, atomic neutral, or molecular.\footnote{For simplicity, we neglect the contribution of ionised molecular species.}
A full account of the efficiency of ejected to accreted material should consider these three states, i.e., $\xi_\mathrm{tot} = \xi_\mathrm{ion} + \xi_\mathrm{ato} + \xi_\mathrm{mol}$ \citep[e.g.,][]{Fedriani2019}.  
Assuming that $\xi_\mathrm{tot}$ remains constant across YSO evolutionary stages, the observed decrease in $\xi_\mathrm{ion}$ with evolutionary stage suggests that jets are predominantly neutral (atomic and molecular) in the earlier protostellar phases, with the ionization fraction increasing as the YSOs evolve. This scenario is supported by previous studies of Herbig-Haro objects HH 34 \citep{Nisini2016} and HH 211 \citep{TomRay2023}.  

\subsection{Accretion and ejection variability}

Regarding the short-term (days to months) time evolution shown in Fig. \ref{fig:fig2}, there is no clear evidence of a temporal correlation between accretion and ionised mass-loss 
rates. This can be understood by considering the timescale of the formation of lobes of ionised gas in the jet, which we approximate as 
$\sim$ 0.5 $\times$ beamsize divided by the jet velocity, or 
$\sim$ $10^2$ days.
\citet{2014AJ....147..125C} reported the variability of near-IR tracers of accretion and winds in a sample of class~{\sc I}
YSOs with a four year survey. They concluded that both types of tracers are sometimes positively or negatively correlated and sometimes uncorrelated, depending on the target. Similar results were obtained by \citet{Ellerbroek2014} in the Herbig Ae/Be star HD 163296 and by \citet{Forbrich} in LRLL 54361. Also, the class 0 YSO HOPS 383, which presented a $\times 35$ increment in bolometric luminosity between 2004 and 2006 \citep{Safron2015}, did not present a corresponding increase in the centimetre flux of the associated radio jet \citep{GM15}. All these results highlight the complexity of the physical relationship between accretion and ejection in jets and winds \citep[see e.g.,][for detailed simulations]{Romanova18,Zhu25}. The observed variability in accretion and ionised mass-loss rates in Fig. \ref{fig:fig2} is probably due to the inherent  stochasticity in these processes \citep{Fischer23_PPVII}.

\subsection{Outliers and complex behaviour} \label{outlier}

The class~{\sc I} YSO CrA43 was excluded from further analysis due to its inferred $\dot{M}_\mathrm{ion}$ being two orders of magnitude larger than the corresponding $\dot{M}_\mathrm{acc}$ (see inset plot of 
Fig. \ref{fig:fig3}). We consider the possible explanation for this anomaly. 
The KMOS observations were taken during a state of quiescence. 
During such an episode of a lower accretion rate, the radio jet core could also become optically thin, allowing non-thermal radio emission from the magnetosphere, if present,  to reach the observer \citep[e.g.,][]{2014ApJ...780..155L}. 
More generally, variability in both accretion \citep[e.g.,][]{Costigan14,Claes22} and jet ejection \citep[e.g.,][]{Reipurth2002,Takami2023} could be an important source of the observed scatter and outliers in studies like the one presented in this paper, even if we attempted to perform observations as simultaneously as possible.

\section{Conclusions} \label{sec:conclusions}

Using multi-epoch coordinated observations with KMOS-VLT and the VLA, we have measured the Br$\gamma$ line and 3.3 cm radio continuum emission 
for 26 YSOs in the Coronet Cluster. Accretion and ionised jet ejection rates were derived from both tracers for 4 YSOs with confirmed detections, while upper limits were obtained for an additional 4 YSOs.  
Using time-averaged measurements, we find a sub-linear relation between $\dot{M}_\mathrm{acc}$ and $\dot{M}_\mathrm{ion}$, with a slope of $\approx$ $0.3 \pm 0.1$ in logarithmic space. This implies that the amount of ionised jet ejection follows that of mass accretion across the protostar to protoplanetary disk evolutionary timescales of up to a few Myr. However, we do not find a clear relationship between these quantities in our time monitoring sampling timescales from days to months, possibly due to the complex interplay between magnetospheric accretion and jet launching processes.   

\section*{Data availability}
The data underlying this article are available in its online supplementary material.

\section*{Acknowledgements}
Based on observations collected at the European Organisation for Astronomical Research in the Southern Hemisphere under ESO programme 093.C-0657.
The National Radio Astronomy Observatory is a facility of the National Science Foundation operated under cooperative agreement by Associated Universities, Inc. 
A.G., R.G.M., J.R.A., and C.C.G. acknowledge support from CONAHCyT Ciencia de Frontera project ID 86372 ``Citlalcóatl''. These authors also 
thank the support from DGAPA-UNAM and UNAM-PAPIIT projects IN105225 and
IG101224. R.G.M. acknowledges support from the ESO Scientific Visitor
Programme. H.B.L. is supported by the National Science and Technology Council (NSTC) of Taiwan (Grant Nos.\ 111-2112-M-110-022-MY3, 113-2112-M-110-022-MY3). The authors thank Stefan Meingast for providing  VISIONS photometric data.
This research has made use of the following software:
\texttt{Astropy} \citep{2018AJ....156..123A, astropy2022}, and
\texttt{CASA} \citep{CASA2022}.




\bibliographystyle{mnras}
\bibliography{Corona_draft} 

\begin{thebibliography}{}
\makeatletter
\relax
\def\mn@urlcharsother{\let\do\@makeother \do\$\do\&\do\#\do\^\do\_\do\%\do\~}
\def\mn@doi{\begingroup\mn@urlcharsother \@ifnextchar [ {\mn@doi@} {\mn@doi@[]}}
\def\mn@doi@[#1]#2{\def\@tempa{#1}\ifx\@tempa\@empty \href {http://dx.doi.org/#2} {doi:#2}\else \href {http://dx.doi.org/#2} {#1}\fi \endgroup}
\def\mn@eprint#1#2{\mn@eprint@#1:#2::\@nil}
\def\mn@eprint@arXiv#1{\href {http://arxiv.org/abs/#1} {{\tt arXiv:#1}}}
\def\mn@eprint@dblp#1{\href {http://dblp.uni-trier.de/rec/bibtex/#1.xml} {dblp:#1}}
\def\mn@eprint@#1:#2:#3:#4\@nil{\def\@tempa {#1}\def\@tempb {#2}\def\@tempc {#3}\ifx \@tempc \@empty \let \@tempc \@tempb \let \@tempb \@tempa \fi \ifx \@tempb \@empty \def\@tempb {arXiv}\fi \@ifundefined {mn@eprint@\@tempb}{\@tempb:\@tempc}{\expandafter \expandafter \csname mn@eprint@\@tempb\endcsname \expandafter{\@tempc}}}

\bibitem[\protect\citeauthoryear{{Adams}, {Lada}  \& {Shu}}{{Adams} et~al.}{1987}]{Adams1987}
{Adams} F.~C.,  {Lada} C.~J.,   {Shu} F.~H.,  1987, \mn@doi [\apj] {10.1086/164924}, \href {https://ui.adsabs.harvard.edu/abs/1987ApJ...312..788A} {312, 788}

\bibitem[\protect\citeauthoryear{{Alcal{\'a}} et~al.,}{{Alcal{\'a}} et~al.}{2017}]{Alcala2017}
{Alcal{\'a}} J.~M.,  et~al., 2017, \mn@doi [\aap] {10.1051/0004-6361/201629929}, \href {https://ui.adsabs.harvard.edu/abs/2017A&A...600A..20A} {600, A20}

\bibitem[\protect\citeauthoryear{{Andre}}{{Andre}}{1987}]{1987Andre}
{Andre} P.,  1987, in {Montmerle} T.,  {Bertout} C.,  eds, Protostars and Molecular Clouds. p.~143

\bibitem[\protect\citeauthoryear{{Andre}, {Ward-Thompson}  \& {Barsony}}{{Andre} et~al.}{1993}]{Andre1993}
{Andre} P.,  {Ward-Thompson} D.,   {Barsony} M.,  1993, \mn@doi [\apj] {10.1086/172425}, \href {https://ui.adsabs.harvard.edu/abs/1993ApJ...406..122A} {406, 122}

\bibitem[\protect\citeauthoryear{{Anglada}, {Rodr{\'\i}guez}  \& {Carrasco-Gonz{\'a}lez}}{{Anglada} et~al.}{2018}]{2018A&ARv..26....3A}
{Anglada} G.,  {Rodr{\'\i}guez} L.~F.,   {Carrasco-Gonz{\'a}lez} C.,  2018, \mn@doi [\aapr] {10.1007/s00159-018-0107-z}, \href {https://ui.adsabs.harvard.edu/abs/2018A&ARv..26....3A} {26, 3}

\bibitem[\protect\citeauthoryear{{Armitage}, {Clarke}  \& {Palla}}{{Armitage} et~al.}{2003}]{Armitage2003}
{Armitage} P.~J.,  {Clarke} C.~J.,   {Palla} F.,  2003, \mn@doi [\mnras] {10.1046/j.1365-8711.2003.06604.x}, \href {https://ui.adsabs.harvard.edu/abs/2003MNRAS.342.1139A} {342, 1139}

\bibitem[\protect\citeauthoryear{{Astropy Collaboration} et~al.,}{{Astropy Collaboration} et~al.}{2018}]{2018AJ....156..123A}
{Astropy Collaboration} et~al., 2018, \mn@doi [\aj] {10.3847/1538-3881/aabc4f}, \href {https://ui.adsabs.harvard.edu/abs/2018AJ....156..123A} {156, 123}

\bibitem[\protect\citeauthoryear{{Astropy Collaboration} et~al.,}{{Astropy Collaboration} et~al.}{2022}]{astropy2022}
{Astropy Collaboration} et~al., 2022, \mn@doi [\apj] {10.3847/1538-4357/ac7c74}, \href {https://ui.adsabs.harvard.edu/abs/2022ApJ...935..167A} {935, 167}

\bibitem[\protect\citeauthoryear{{Bacciotti} et~al.,}{{Bacciotti} et~al.}{2025}]{Bacciotti2025}
{Bacciotti} F.,  et~al., 2025, \mn@doi [arXiv e-prints] {10.48550/arXiv.2501.03920}, \href {https://ui.adsabs.harvard.edu/abs/2025arXiv250103920B} {p. arXiv:2501.03920}

\bibitem[\protect\citeauthoryear{{Banzatti}, {Pascucci}, {Edwards}, {Fang}, {Gorti}  \& {Flock}}{{Banzatti} et~al.}{2019}]{Banzatti2019}
{Banzatti} A.,  {Pascucci} I.,  {Edwards} S.,  {Fang} M.,  {Gorti} U.,   {Flock} M.,  2019, \mn@doi [\apj] {10.3847/1538-4357/aaf1aa}, \href {https://ui.adsabs.harvard.edu/abs/2019ApJ...870...76B} {870, 76}

\bibitem[\protect\citeauthoryear{{CASA Team} et~al.,}{{CASA Team} et~al.}{2022}]{CASA2022}
{CASA Team} et~al., 2022, \mn@doi [\pasp] {10.1088/1538-3873/ac9642}, \href {https://ui.adsabs.harvard.edu/abs/2022PASP..134k4501C} {134, 114501}

\bibitem[\protect\citeauthoryear{{Calvet} \& {Gullbring}}{{Calvet} \& {Gullbring}}{1998}]{CalvetGullbring98}
{Calvet} N.,  {Gullbring} E.,  1998, \mn@doi [\apj] {10.1086/306527}, \href {https://ui.adsabs.harvard.edu/abs/1998ApJ...509..802C} {509, 802}

\bibitem[\protect\citeauthoryear{{Cazzoletti} et~al.,}{{Cazzoletti} et~al.}{2019}]{Cazzoletti2019}
{Cazzoletti} P.,  et~al., 2019, \mn@doi [\aap] {10.1051/0004-6361/201935273}, \href {https://ui.adsabs.harvard.edu/abs/2019A&A...626A..11C} {626, A11}

\bibitem[\protect\citeauthoryear{{Choi}, {Hamaguchi}, {Lee}  \& {Tatematsu}}{{Choi} et~al.}{2008}]{Choi2008}
{Choi} M.,  {Hamaguchi} K.,  {Lee} J.-E.,   {Tatematsu} K.,  2008, \mn@doi [\apj] {10.1086/591540}, \href {https://ui.adsabs.harvard.edu/abs/2008ApJ...687..406C} {687, 406}

\bibitem[\protect\citeauthoryear{{Claes} et~al.,}{{Claes} et~al.}{2022}]{Claes22}
{Claes} R.~A.~B.,  et~al., 2022, \mn@doi [\aap] {10.1051/0004-6361/202244135}, \href {https://ui.adsabs.harvard.edu/abs/2022A&A...664L...7C} {664, L7}

\bibitem[\protect\citeauthoryear{{Connelley} \& {Greene}}{{Connelley} \& {Greene}}{2014}]{2014AJ....147..125C}
{Connelley} M.~S.,  {Greene} T.~P.,  2014, \mn@doi [\aj] {10.1088/0004-6256/147/6/125}, \href {https://ui.adsabs.harvard.edu/abs/2014AJ....147..125C} {147, 125}

\bibitem[\protect\citeauthoryear{{Contreras Pe{\~n}a} et~al.,}{{Contreras Pe{\~n}a} et~al.}{2017}]{2017MNRAS.465.3039C}
{Contreras Pe{\~n}a} C.,  et~al., 2017, \mn@doi [\mnras] {10.1093/mnras/stw2802}, \href {https://ui.adsabs.harvard.edu/abs/2017MNRAS.465.3039C} {465, 3039}

\bibitem[\protect\citeauthoryear{{Costigan}, {Vink}, {Scholz}, {Ray}  \& {Testi}}{{Costigan} et~al.}{2014}]{Costigan14}
{Costigan} G.,  {Vink} J.~S.,  {Scholz} A.,  {Ray} T.,   {Testi} L.,  2014, \mn@doi [\mnras] {10.1093/mnras/stu529}, \href {https://ui.adsabs.harvard.edu/abs/2014MNRAS.440.3444C} {440, 3444}

\bibitem[\protect\citeauthoryear{{Davies} et~al.,}{{Davies} et~al.}{2013}]{2013A&A...558A..56D}
{Davies} R.~I.,  et~al., 2013, \mn@doi [\aap] {10.1051/0004-6361/201322282}, \href {https://ui.adsabs.harvard.edu/abs/2013A&A...558A..56D} {558, A56}

\bibitem[\protect\citeauthoryear{{Deller}, {Forbrich}  \& {Loinard}}{{Deller} et~al.}{2013}]{2013Forbrich}
{Deller} A.~T.,  {Forbrich} J.,   {Loinard} L.,  2013, \mn@doi [\aap] {10.1051/0004-6361/201321085}, \href {https://ui.adsabs.harvard.edu/abs/2013A&A...552A..51D} {552, A51}

\bibitem[\protect\citeauthoryear{{Dong}, {Najita}  \& {Brittain}}{{Dong} et~al.}{2018}]{Dong2018}
{Dong} R.,  {Najita} J.~R.,   {Brittain} S.,  2018, \mn@doi [\apj] {10.3847/1538-4357/aaccfc}, \href {https://ui.adsabs.harvard.edu/abs/2018ApJ...862..103D} {862, 103}

\bibitem[\protect\citeauthoryear{{Dunham} et~al.,}{{Dunham} et~al.}{2014}]{Dunham14_PPVI}
{Dunham} M.~M.,  et~al., 2014, in {Beuther} H.,  {Klessen} R.~S.,  {Dullemond} C.~P.,   {Henning} T.,  eds, Protostars and Planets VI. pp 195--218 (\mn@eprint {arXiv} {1401.1809}), \mn@doi{10.2458/azu_uapress_9780816531240-ch009}

\bibitem[\protect\citeauthoryear{{Dzib}, {Loinard}, {Ortiz-Le{\'o}n}, {Rodr{\'\i}guez}  \& {Galli}}{{Dzib} et~al.}{2018}]{Dzib2018}
{Dzib} S.~A.,  {Loinard} L.,  {Ortiz-Le{\'o}n} G.~N.,  {Rodr{\'\i}guez} L.~F.,   {Galli} P. A.~B.,  2018, \mn@doi [\apj] {10.3847/1538-4357/aae687}, \href {https://ui.adsabs.harvard.edu/abs/2018ApJ...867..151D} {867, 151}

\bibitem[\protect\citeauthoryear{{Ellerbroek} et~al.,}{{Ellerbroek} et~al.}{2014}]{Ellerbroek2014}
{Ellerbroek} L.~E.,  et~al., 2014, \mn@doi [\aap] {10.1051/0004-6361/201323092}, \href {https://ui.adsabs.harvard.edu/abs/2014A&A...563A..87E} {563, A87}

\bibitem[\protect\citeauthoryear{{Fang} et~al.,}{{Fang} et~al.}{2018}]{Fang2018}
{Fang} M.,  et~al., 2018, \mn@doi [\apj] {10.3847/1538-4357/aae780}, \href {https://ui.adsabs.harvard.edu/abs/2018ApJ...868...28F} {868, 28}

\bibitem[\protect\citeauthoryear{{Fedriani} et~al.,}{{Fedriani} et~al.}{2019}]{Fedriani2019}
{Fedriani} R.,  et~al., 2019, \mn@doi [Nature Communications] {10.1038/s41467-019-11595-x}, \href {https://ui.adsabs.harvard.edu/abs/2019NatCo..10.3630F} {10, 3630}

\bibitem[\protect\citeauthoryear{{Feigelson} \& {Montmerle}}{{Feigelson} \& {Montmerle}}{1999}]{Feigelson99ARAA}
{Feigelson} E.~D.,  {Montmerle} T.,  1999, \mn@doi [\araa] {10.1146/annurev.astro.37.1.363}, \href {https://ui.adsabs.harvard.edu/abs/1999ARA&A..37..363F} {37, 363}

\bibitem[\protect\citeauthoryear{{Fiorellino} et~al.,}{{Fiorellino} et~al.}{2021}]{2021A&A...650A..43F}
{Fiorellino} E.,  et~al., 2021, \mn@doi [\aap] {10.1051/0004-6361/202039264}, \href {https://ui.adsabs.harvard.edu/abs/2021A&A...650A..43F} {650, A43}

\bibitem[\protect\citeauthoryear{{Fischer}, {Hillenbrand}, {Herczeg}, {Johnstone}, {Kospal}  \& {Dunham}}{{Fischer} et~al.}{2023}]{Fischer23_PPVII}
{Fischer} W.~J.,  {Hillenbrand} L.~A.,  {Herczeg} G.~J.,  {Johnstone} D.,  {Kospal} A.,   {Dunham} M.~M.,  2023, in {Inutsuka} S.,  {Aikawa} Y.,  {Muto} T.,  {Tomida} K.,   {Tamura} M.,  eds,  Astronomical Society of the Pacific Conference Series Vol. 534, Protostars and Planets VII. p.~355 (\mn@eprint {arXiv} {2203.11257}), \mn@doi{10.48550/arXiv.2203.11257}

\bibitem[\protect\citeauthoryear{{Folha} \& {Emerson}}{{Folha} \& {Emerson}}{2001}]{FolhaEmerson2001}
{Folha} D.~F.~M.,  {Emerson} J.~P.,  2001, \mn@doi [\aap] {10.1051/0004-6361:20000018}, \href {https://ui.adsabs.harvard.edu/abs/2001A&A...365...90F} {365, 90}

\bibitem[\protect\citeauthoryear{{Forbrich} \& {Preibisch}}{{Forbrich} \& {Preibisch}}{2007}]{2007A&A...475..959F}
{Forbrich} J.,  {Preibisch} T.,  2007, \mn@doi [\aap] {10.1051/0004-6361:20066342}, \href {https://ui.adsabs.harvard.edu/abs/2007A&A...475..959F} {475, 959}

\bibitem[\protect\citeauthoryear{{Forbrich} et~al.,}{{Forbrich} et~al.}{2007}]{Forbrich2007}
{Forbrich} J.,  et~al., 2007, \mn@doi [\aap] {10.1051/0004-6361:20066158}, \href {https://ui.adsabs.harvard.edu/abs/2007A&A...464.1003F} {464, 1003}

\bibitem[\protect\citeauthoryear{{Forbrich}, {Rodr{\'\i}guez}, {Palau}, {Zapata}, {Muzerolle}  \& {Gutermuth}}{{Forbrich} et~al.}{2015}]{Forbrich}
{Forbrich} J.,  {Rodr{\'\i}guez} L.~F.,  {Palau} A.,  {Zapata} L.~A.,  {Muzerolle} J.,   {Gutermuth} R.~A.,  2015, \mn@doi [\apj] {10.1088/0004-637X/814/1/15}, \href {https://ui.adsabs.harvard.edu/abs/2015ApJ...814...15F} {814, 15}

\bibitem[\protect\citeauthoryear{{Forbrich}, {Dzib}, {Reid}  \& {Menten}}{{Forbrich} et~al.}{2021}]{Forbrich21}
{Forbrich} J.,  {Dzib} S.~A.,  {Reid} M.~J.,   {Menten} K.~M.,  2021, \mn@doi [\apj] {10.3847/1538-4357/abc68e}, \href {https://ui.adsabs.harvard.edu/abs/2021ApJ...906...23F} {906, 23}

\bibitem[\protect\citeauthoryear{{Frank} et~al.,}{{Frank} et~al.}{2014}]{Frank14_PPVI}
{Frank} A.,  et~al., 2014, in {Beuther} H.,  {Klessen} R.~S.,  {Dullemond} C.~P.,   {Henning} T.,  eds, Protostars and Planets VI. pp 451--474 (\mn@eprint {arXiv} {1402.3553}), \mn@doi{10.2458/azu_uapress_9780816531240-ch020}

\bibitem[\protect\citeauthoryear{{Freudling}, {Romaniello}, {Bramich}, {Ballester}, {Forchi}, {Garc{\'\i}a-Dabl{\'o}}, {Moehler}  \& {Neeser}}{{Freudling} et~al.}{2013}]{2013A&A...559A..96F}
{Freudling} W.,  {Romaniello} M.,  {Bramich} D.~M.,  {Ballester} P.,  {Forchi} V.,  {Garc{\'\i}a-Dabl{\'o}} C.~E.,  {Moehler} S.,   {Neeser} M.~J.,  2013, \mn@doi [\aap] {10.1051/0004-6361/201322494}, \href {https://ui.adsabs.harvard.edu/abs/2013A&A...559A..96F} {559, A96}

\bibitem[\protect\citeauthoryear{{Gaidos}, {Gehrig}  \& {G{\"u}del}}{{Gaidos} et~al.}{2025}]{Gaidos2025}
{Gaidos} E.,  {Gehrig} L.,   {G{\"u}del} M.,  2025, \mn@doi [\aap] {10.1051/0004-6361/202453207}, \href {https://ui.adsabs.harvard.edu/abs/2025A&A...696A.207G} {696, A207}

\bibitem[\protect\citeauthoryear{{Galli}, {Bouy}, {Olivares}, {Miret-Roig}, {Sarro}, {Barrado}, {Berihuete}  \& {Brandner}}{{Galli} et~al.}{2020}]{Galli2020}
{Galli} P.~A.~B.,  {Bouy} H.,  {Olivares} J.,  {Miret-Roig} N.,  {Sarro} L.~M.,  {Barrado} D.,  {Berihuete} A.,   {Brandner} W.,  2020, \mn@doi [\aap] {10.1051/0004-6361/201936708}, \href {https://ui.adsabs.harvard.edu/abs/2020A&A...634A..98G} {634, A98}

\bibitem[\protect\citeauthoryear{{Galv{\'a}n-Madrid} et~al.,}{{Galv{\'a}n-Madrid} et~al.}{2014}]{GM2014}
{Galv{\'a}n-Madrid} R.,  et~al., 2014, \mn@doi [\aap] {10.1051/0004-6361/201424630}, \href {https://ui.adsabs.harvard.edu/abs/2014A&A...570L...9G} {570, L9}

\bibitem[\protect\citeauthoryear{{Galv{\'a}n-Madrid}, {Rodr{\'\i}guez}, {Liu}, {Costigan}, {Palau}, {Zapata}  \& {Loinard}}{{Galv{\'a}n-Madrid} et~al.}{2015}]{GM15}
{Galv{\'a}n-Madrid} R.,  {Rodr{\'\i}guez} L.~F.,  {Liu} H.~B.,  {Costigan} G.,  {Palau} A.,  {Zapata} L.~A.,   {Loinard} L.,  2015, \mn@doi [\apjl] {10.1088/2041-8205/806/2/L32}, \href {https://ui.adsabs.harvard.edu/abs/2015ApJ...806L..32G} {806, L32}

\bibitem[\protect\citeauthoryear{{Garufi} et~al.,}{{Garufi} et~al.}{2025}]{Garufi2025}
{Garufi} A.,  et~al., 2025, \mn@doi [\aap] {10.1051/0004-6361/202452496}, \href {https://ui.adsabs.harvard.edu/abs/2025A&A...694A.290G} {694, A290}

\bibitem[\protect\citeauthoryear{{Guo} et~al.,}{{Guo} et~al.}{2020}]{2020MNRAS.492..294G}
{Guo} Z.,  et~al., 2020, \mn@doi [\mnras] {10.1093/mnras/stz3374}, \href {https://ui.adsabs.harvard.edu/abs/2020MNRAS.492..294G} {492, 294}

\bibitem[\protect\citeauthoryear{{Hartigan}, {Edwards}  \& {Ghandour}}{{Hartigan} et~al.}{1995}]{Hartigan1995}
{Hartigan} P.,  {Edwards} S.,   {Ghandour} L.,  1995, \mn@doi [\apj] {10.1086/176344}, \href {https://ui.adsabs.harvard.edu/abs/1995ApJ...452..736H} {452, 736}

\bibitem[\protect\citeauthoryear{{Hartmann}, {Herczeg}  \& {Calvet}}{{Hartmann} et~al.}{2016}]{Hartmann2016}
{Hartmann} L.,  {Herczeg} G.,   {Calvet} N.,  2016, \mn@doi [\araa] {10.1146/annurev-astro-081915-023347}, \href {https://ui.adsabs.harvard.edu/abs/2016ARA&A..54..135H} {54, 135}

\bibitem[\protect\citeauthoryear{{Hsieh}, {Arce}, {Maureira}, {Pineda}, {Segura-Cox}, {Mardones}, {Dunham}  \& {Arun}}{{Hsieh} et~al.}{2024}]{Hsieh24}
{Hsieh} C.-H.,  {Arce} H.~G.,  {Maureira} M.~J.,  {Pineda} J.~E.,  {Segura-Cox} D.,  {Mardones} D.,  {Dunham} M.~M.,   {Arun} A.,  2024, \mn@doi [\apj] {10.3847/1538-4357/ad6152}, \href {https://ui.adsabs.harvard.edu/abs/2024ApJ...973..138H} {973, 138}

\bibitem[\protect\citeauthoryear{{Liu} et~al.,}{{Liu} et~al.}{2014}]{2014ApJ...780..155L}
{Liu} H.~B.,  et~al., 2014, \mn@doi [\apj] {10.1088/0004-637X/780/2/155}, \href {https://ui.adsabs.harvard.edu/abs/2014ApJ...780..155L} {780, 155}

\bibitem[\protect\citeauthoryear{{Lora}, {Nony}, {Esquivel}  \& {Galv{\'a}n-Madrid}}{{Lora} et~al.}{2024}]{Lora24}
{Lora} V.,  {Nony} T.,  {Esquivel} A.,   {Galv{\'a}n-Madrid} R.,  2024, \mn@doi [\apj] {10.3847/1538-4357/ad13ed}, \href {https://ui.adsabs.harvard.edu/abs/2024ApJ...962...66L} {962, 66}

\bibitem[\protect\citeauthoryear{{Mac{\'\i}as} et~al.,}{{Mac{\'\i}as} et~al.}{2016}]{Macias2016}
{Mac{\'\i}as} E.,  et~al., 2016, \mn@doi [\apj] {10.3847/0004-637X/829/1/1}, \href {https://ui.adsabs.harvard.edu/abs/2016ApJ...829....1M} {829, 1}

\bibitem[\protect\citeauthoryear{{Manara}, {Ansdell}, {Rosotti}, {Hughes}, {Armitage}, {Lodato}  \& {Williams}}{{Manara} et~al.}{2023}]{Manara23_PPVII}
{Manara} C.~F.,  {Ansdell} M.,  {Rosotti} G.~P.,  {Hughes} A.~M.,  {Armitage} P.~J.,  {Lodato} G.,   {Williams} J.~P.,  2023, in {Inutsuka} S.,  {Aikawa} Y.,  {Muto} T.,  {Tomida} K.,   {Tamura} M.,  eds,  Astronomical Society of the Pacific Conference Series Vol. 534, Protostars and Planets VII. p.~539 (\mn@eprint {arXiv} {2203.09930}), \mn@doi{10.48550/arXiv.2203.09930}

\bibitem[\protect\citeauthoryear{{Meingast} et~al.,}{{Meingast} et~al.}{2023}]{VISIONS}
{Meingast} S.,  et~al., 2023, \mn@doi [\aap] {10.1051/0004-6361/202245771}, \href {https://ui.adsabs.harvard.edu/abs/2023A&A...673A..58M} {673, A58}

\bibitem[\protect\citeauthoryear{{Miettinen}, {Kontinen}, {Harju}  \& {Higdon}}{{Miettinen} et~al.}{2008}]{2008A&A...486..799M}
{Miettinen} O.,  {Kontinen} S.,  {Harju} J.,   {Higdon} J.~L.,  2008, \mn@doi [\aap] {10.1051/0004-6361:200809348}, \href {https://ui.adsabs.harvard.edu/abs/2008A&A...486..799M} {486, 799}

\bibitem[\protect\citeauthoryear{{Morales-Calder{\'o}n} et~al.,}{{Morales-Calder{\'o}n} et~al.}{2011}]{MoralesCalderon2011}
{Morales-Calder{\'o}n} M.,  et~al., 2011, \mn@doi [\apj] {10.1088/0004-637X/733/1/50}, \href {https://ui.adsabs.harvard.edu/abs/2011ApJ...733...50M} {733, 50}

\bibitem[\protect\citeauthoryear{{Muzerolle}, {Calvet}  \& {Hartmann}}{{Muzerolle} et~al.}{2001}]{Muzerolle2001}
{Muzerolle} J.,  {Calvet} N.,   {Hartmann} L.,  2001, \mn@doi [\apj] {10.1086/319779}, \href {https://ui.adsabs.harvard.edu/abs/2001ApJ...550..944M} {550, 944}

\bibitem[\protect\citeauthoryear{{Myers}}{{Myers}}{2009}]{2009ApJ...700.1609M}
{Myers} P.~C.,  2009, \mn@doi [\apj] {10.1088/0004-637X/700/2/1609}, \href {https://ui.adsabs.harvard.edu/abs/2009ApJ...700.1609M} {700, 1609}

\bibitem[\protect\citeauthoryear{{Nisini}, {Antoniucci}, {Giannini}  \& {Lorenzetti}}{{Nisini} et~al.}{2005a}]{2005A&A...429..543N}
{Nisini} B.,  {Antoniucci} S.,  {Giannini} T.,   {Lorenzetti} D.,  2005a, \mn@doi [\aap] {10.1051/0004-6361:20041409}, \href {https://ui.adsabs.harvard.edu/abs/2005A&A...429..543N} {429, 543}

\bibitem[\protect\citeauthoryear{{Nisini}, {Antoniucci}, {Giannini}  \& {Lorenzetti}}{{Nisini} et~al.}{2005b}]{Nisini2005}
{Nisini} B.,  {Antoniucci} S.,  {Giannini} T.,   {Lorenzetti} D.,  2005b, \mn@doi [\aap] {10.1051/0004-6361:20041409}, \href {https://ui.adsabs.harvard.edu/abs/2005A&A...429..543N} {429, 543}

\bibitem[\protect\citeauthoryear{{Nisini}, {Giannini}, {Antoniucci}, {Alcal{\'a}}, {Bacciotti}  \& {Podio}}{{Nisini} et~al.}{2016}]{Nisini2016}
{Nisini} B.,  {Giannini} T.,  {Antoniucci} S.,  {Alcal{\'a}} J.~M.,  {Bacciotti} F.,   {Podio} L.,  2016, \mn@doi [\aap] {10.1051/0004-6361/201628853}, \href {https://ui.adsabs.harvard.edu/abs/2016A&A...595A..76N} {595, A76}

\bibitem[\protect\citeauthoryear{{Nisini}, {Antoniucci}, {Alcal{\'a}}, {Giannini}, {Manara}, {Natta}, {Fedele}  \& {Biazzo}}{{Nisini} et~al.}{2018}]{Nisini2018}
{Nisini} B.,  {Antoniucci} S.,  {Alcal{\'a}} J.~M.,  {Giannini} T.,  {Manara} C.~F.,  {Natta} A.,  {Fedele} D.,   {Biazzo} K.,  2018, \mn@doi [\aap] {10.1051/0004-6361/201730834}, \href {https://ui.adsabs.harvard.edu/abs/2018A&A...609A..87N} {609, A87}

\bibitem[\protect\citeauthoryear{{Nisini} et~al.,}{{Nisini} et~al.}{2024}]{Nisini2024}
{Nisini} B.,  et~al., 2024, \mn@doi [\apj] {10.3847/1538-4357/ad3d5a}, \href {https://ui.adsabs.harvard.edu/abs/2024ApJ...967..168N} {967, 168}

\bibitem[\protect\citeauthoryear{{Pascucci}, {Ricci}, {Gorti}, {Hollenbach}, {Hendler}, {Brooks}  \& {Contreras}}{{Pascucci} et~al.}{2014}]{Pascucci2014}
{Pascucci} I.,  {Ricci} L.,  {Gorti} U.,  {Hollenbach} D.,  {Hendler} N.~P.,  {Brooks} K.~J.,   {Contreras} Y.,  2014, \mn@doi [\apj] {10.1088/0004-637X/795/1/1}, \href {https://ui.adsabs.harvard.edu/abs/2014ApJ...795....1P} {795, 1}

\bibitem[\protect\citeauthoryear{{Pascucci}, {Cabrit}, {Edwards}, {Gorti}, {Gressel}  \& {Suzuki}}{{Pascucci} et~al.}{2023}]{Pascucci2023}
{Pascucci} I.,  {Cabrit} S.,  {Edwards} S.,  {Gorti} U.,  {Gressel} O.,   {Suzuki} T.~K.,  2023, in {Inutsuka} S.,  {Aikawa} Y.,  {Muto} T.,  {Tomida} K.,   {Tamura} M.,  eds,  Astronomical Society of the Pacific Conference Series Vol. 534, Protostars and Planets VII. p.~567 (\mn@eprint {arXiv} {2203.10068}), \mn@doi{10.48550/arXiv.2203.10068}

\bibitem[\protect\citeauthoryear{{Pelletier} \& {Pudritz}}{{Pelletier} \& {Pudritz}}{1992}]{Pelletier1992}
{Pelletier} G.,  {Pudritz} R.~E.,  1992, \mn@doi [\apj] {10.1086/171565}, \href {https://ui.adsabs.harvard.edu/abs/1992ApJ...394..117P} {394, 117}

\bibitem[\protect\citeauthoryear{{Peterson} et~al.,}{{Peterson} et~al.}{2011}]{2011ApJS..194...43P}
{Peterson} D.~E.,  et~al., 2011, \mn@doi [\apjs] {10.1088/0067-0049/194/2/43}, \href {https://ui.adsabs.harvard.edu/abs/2011ApJS..194...43P} {194, 43}

\bibitem[\protect\citeauthoryear{{Pudritz}, {Ouyed}, {Fendt}  \& {Brandenburg}}{{Pudritz} et~al.}{2007}]{Pudritz05_PPV}
{Pudritz} R.~E.,  {Ouyed} R.,  {Fendt} C.,   {Brandenburg} A.,  2007, in {Reipurth} B.,  {Jewitt} D.,   {Keil} K.,  eds, Protostars and Planets V. p.~277 (\mn@eprint {arXiv} {astro-ph/0603592}), \mn@doi{10.48550/arXiv.astro-ph/0603592}

\bibitem[\protect\citeauthoryear{{Ray} et~al.,}{{Ray} et~al.}{2023}]{TomRay2023}
{Ray} T.~P.,  et~al., 2023, \mn@doi [\nat] {10.1038/s41586-023-06551-1}, \href {https://ui.adsabs.harvard.edu/abs/2023Natur.622...48R} {622, 48}

\bibitem[\protect\citeauthoryear{{Reipurth}, {Rodr{\'\i}guez}, {Anglada}  \& {Bally}}{{Reipurth} et~al.}{2002}]{Reipurth2002}
{Reipurth} B.,  {Rodr{\'\i}guez} L.~F.,  {Anglada} G.,   {Bally} J.,  2002, \mn@doi [\aj] {10.1086/341172}, \href {https://ui.adsabs.harvard.edu/abs/2002AJ....124.1045R} {124, 1045}

\bibitem[\protect\citeauthoryear{{Robitaille}, {Whitney}, {Indebetouw}  \& {Wood}}{{Robitaille} et~al.}{2007}]{Robitaille}
{Robitaille} T.~P.,  {Whitney} B.~A.,  {Indebetouw} R.,   {Wood} K.,  2007, \mn@doi [\apjs] {10.1086/512039}, \href {https://ui.adsabs.harvard.edu/abs/2007ApJS..169..328R} {169, 328}

\bibitem[\protect\citeauthoryear{{Rodr{\'\i}guez}, {Zapata}, {Dzib}, {Ortiz-Le{\'o}n}, {Loinard}, {Mac{\'\i}as}  \& {Anglada}}{{Rodr{\'\i}guez} et~al.}{2014}]{Rodriguez2014}
{Rodr{\'\i}guez} L.~F.,  {Zapata} L.~A.,  {Dzib} S.~A.,  {Ortiz-Le{\'o}n} G.~N.,  {Loinard} L.,  {Mac{\'\i}as} E.,   {Anglada} G.,  2014, \mn@doi [\apjl] {10.1088/2041-8205/793/1/L21}, \href {https://ui.adsabs.harvard.edu/abs/2014ApJ...793L..21R} {793, L21}

\bibitem[\protect\citeauthoryear{{Romanova}, {Blinova}, {Ustyugova}, {Koldoba}  \& {Lovelace}}{{Romanova} et~al.}{2018}]{Romanova18}
{Romanova} M.~M.,  {Blinova} A.~A.,  {Ustyugova} G.~V.,  {Koldoba} A.~V.,   {Lovelace} R.~V.~E.,  2018, \mn@doi [\na] {10.1016/j.newast.2018.01.011}, \href {https://ui.adsabs.harvard.edu/abs/2018NewA...62...94R} {62, 94}

\bibitem[\protect\citeauthoryear{{Rota}, {Meijerhof}, {van der Marel}, {Francis}, {van der Tak}  \& {Sellek}}{{Rota} et~al.}{2024}]{Rota2024}
{Rota} A.~A.,  {Meijerhof} J.~D.,  {van der Marel} N.,  {Francis} L.,  {van der Tak} F.~F.~S.,   {Sellek} A.~D.,  2024, \mn@doi [\aap] {10.1051/0004-6361/202348387}, \href {https://ui.adsabs.harvard.edu/abs/2024A&A...684A.134R} {684, A134}

\bibitem[\protect\citeauthoryear{{Rota} et~al.,}{{Rota} et~al.}{2025}]{Rota2025}
{Rota} A.~A.,  et~al., 2025, \mn@doi [arXiv e-prints] {10.48550/arXiv.2505.16586}, \href {https://ui.adsabs.harvard.edu/abs/2025arXiv250516586R} {p. arXiv:2505.16586}

\bibitem[\protect\citeauthoryear{{Safron} et~al.,}{{Safron} et~al.}{2015}]{Safron2015}
{Safron} E.~J.,  et~al., 2015, \mn@doi [\apjl] {10.1088/2041-8205/800/1/L5}, \href {https://ui.adsabs.harvard.edu/abs/2015ApJ...800L...5S} {800, L5}

\bibitem[\protect\citeauthoryear{{Sharples} et~al.,}{{Sharples} et~al.}{2013}]{Sharples2013}
{Sharples} R.,  et~al., 2013, The Messenger, \href {https://ui.adsabs.harvard.edu/abs/2013Msngr.151...21S} {151, 21}

\bibitem[\protect\citeauthoryear{{Sheehan}, {Tobin}, {Federman}, {Megeath}  \& {Looney}}{{Sheehan} et~al.}{2020}]{Sheehan2020}
{Sheehan} P.~D.,  {Tobin} J.~J.,  {Federman} S.,  {Megeath} S.~T.,   {Looney} L.~W.,  2020, \mn@doi [\apj] {10.3847/1538-4357/abbad5}, \href {https://ui.adsabs.harvard.edu/abs/2020ApJ...902..141S} {902, 141}

\bibitem[\protect\citeauthoryear{{Shu}, {Najita}, {Ostriker}, {Wilkin}, {Ruden}  \& {Lizano}}{{Shu} et~al.}{1994}]{Shu1994}
{Shu} F.,  {Najita} J.,  {Ostriker} E.,  {Wilkin} F.,  {Ruden} S.,   {Lizano} S.,  1994, \mn@doi [\apj] {10.1086/174363}, \href {https://ui.adsabs.harvard.edu/abs/1994ApJ...429..781S} {429, 781}

\bibitem[\protect\citeauthoryear{{Sicilia-Aguilar}, {Henning}, {Juh{\'a}sz}, {Bouwman}, {Garmire}  \& {Garmire}}{{Sicilia-Aguilar} et~al.}{2008}]{Aurora2008}
{Sicilia-Aguilar} A.,  {Henning} T.,  {Juh{\'a}sz} A.,  {Bouwman} J.,  {Garmire} G.,   {Garmire} A.,  2008, \mn@doi [\apj] {10.1086/591932}, \href {https://ui.adsabs.harvard.edu/abs/2008ApJ...687.1145S} {687, 1145}

\bibitem[\protect\citeauthoryear{{Sicilia-Aguilar}, {Henning}, {Kainulainen}  \& {Roccatagliata}}{{Sicilia-Aguilar} et~al.}{2011}]{2011ApJ...736..137S}
{Sicilia-Aguilar} A.,  {Henning} T.,  {Kainulainen} J.,   {Roccatagliata} V.,  2011, \mn@doi [\apj] {10.1088/0004-637X/736/2/137}, \href {https://ui.adsabs.harvard.edu/abs/2011ApJ...736..137S} {736, 137}

\bibitem[\protect\citeauthoryear{{Skrutskie} et~al.,}{{Skrutskie} et~al.}{2006}]{2MASS2006}
{Skrutskie} M.~F.,  et~al., 2006, \mn@doi [\aj] {10.1086/498708}, \href {https://ui.adsabs.harvard.edu/abs/2006AJ....131.1163S} {131, 1163}

\bibitem[\protect\citeauthoryear{{Takami} et~al.,}{{Takami} et~al.}{2006}]{2006ApJ...641..357T}
{Takami} M.,  et~al., 2006, \mn@doi [\apj] {10.1086/500352}, \href {https://ui.adsabs.harvard.edu/abs/2006ApJ...641..357T} {641, 357}

\bibitem[\protect\citeauthoryear{{Takami} et~al.,}{{Takami} et~al.}{2020}]{Takami2020}
{Takami} M.,  et~al., 2020, \mn@doi [\apj] {10.3847/1538-4357/abab98}, \href {https://ui.adsabs.harvard.edu/abs/2020ApJ...901...24T} {901, 24}

\bibitem[\protect\citeauthoryear{{Takami} et~al.,}{{Takami} et~al.}{2023}]{Takami2023}
{Takami} M.,  et~al., 2023, \mn@doi [\apjs] {10.3847/1538-4365/ac9afc}, \href {https://ui.adsabs.harvard.edu/abs/2023ApJS..264....1T} {264, 1}

\bibitem[\protect\citeauthoryear{{Testi} et~al.,}{{Testi} et~al.}{2025}]{Testi2025}
{Testi} L.,  et~al., 2025, \mn@doi [\aap] {10.1051/0004-6361/202554149}, \href {https://ui.adsabs.harvard.edu/abs/2025A&A...703A.277T} {703, A277}

\bibitem[\protect\citeauthoryear{{Tychoniec} et~al.,}{{Tychoniec} et~al.}{2018}]{Tychoniec2018}
{Tychoniec} {\L}.,  et~al., 2018, \mn@doi [\apjs] {10.3847/1538-4365/aaceae}, \href {https://ui.adsabs.harvard.edu/abs/2018ApJS..238...19T} {238, 19}

\bibitem[\protect\citeauthoryear{{Williams} \& {Cieza}}{{Williams} \& {Cieza}}{2011}]{WillamsCieza2011}
{Williams} J.~P.,  {Cieza} L.~A.,  2011, \mn@doi [\araa] {10.1146/annurev-astro-081710-102548}, \href {https://ui.adsabs.harvard.edu/abs/2011ARA&A..49...67W} {49, 67}

\bibitem[\protect\citeauthoryear{{Zhu}}{{Zhu}}{2025}]{Zhu25}
{Zhu} Z.,  2025, \mn@doi [\mnras] {10.1093/mnras/staf250}, \href {https://ui.adsabs.harvard.edu/abs/2025MNRAS.537.3701Z} {537, 3701}

\makeatother
\end{thebibliography}

\vspace{5mm}





\appendix
\section*{Supplementary Material}
\begin{figure*}
    \centering
    \begin{subfigure}[b]{0.22\textwidth}
        \includegraphics[width=\textwidth]{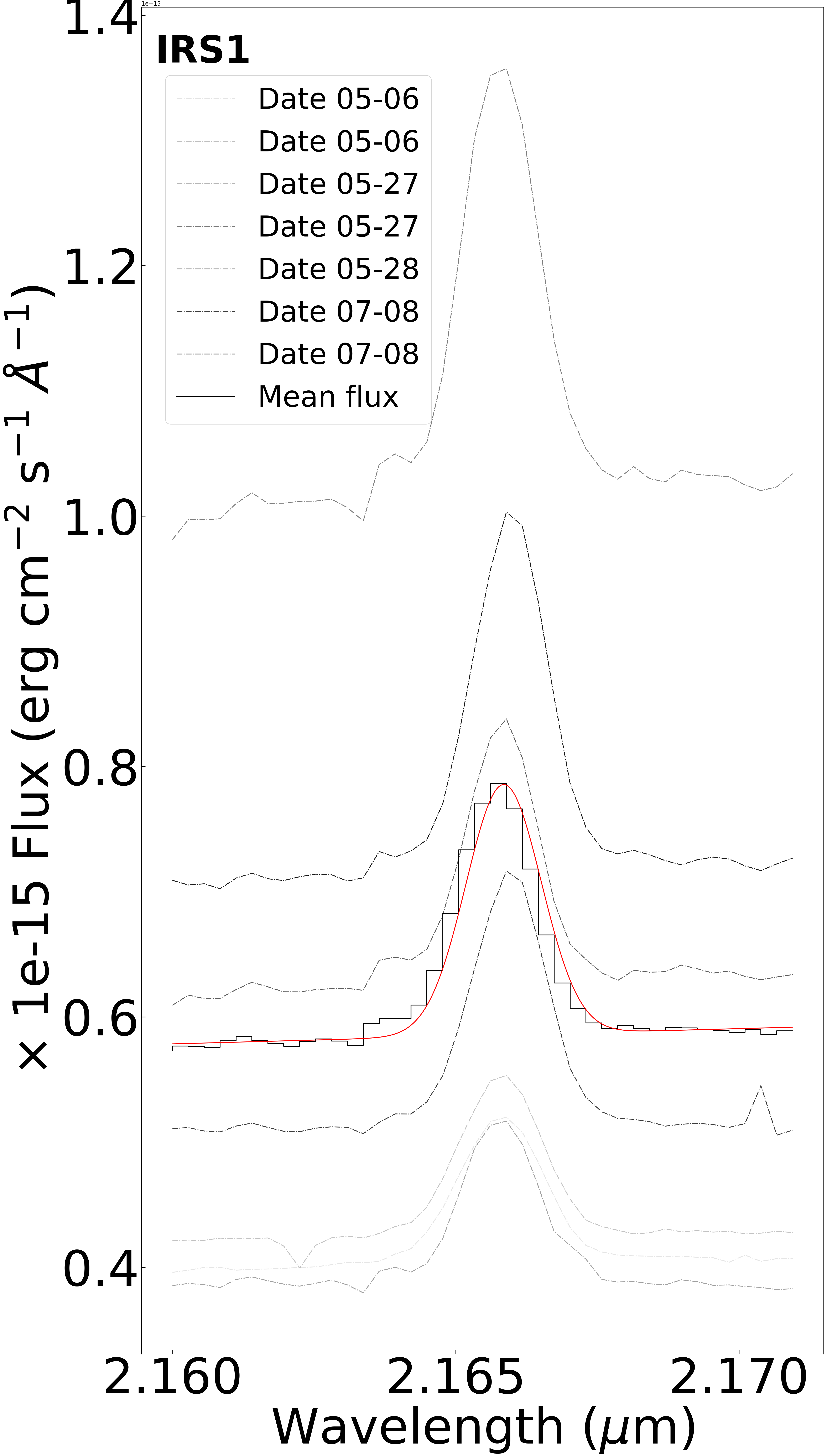}
    \end{subfigure}
    \begin{subfigure}[b]{0.22\textwidth}
        \includegraphics[width=\textwidth]{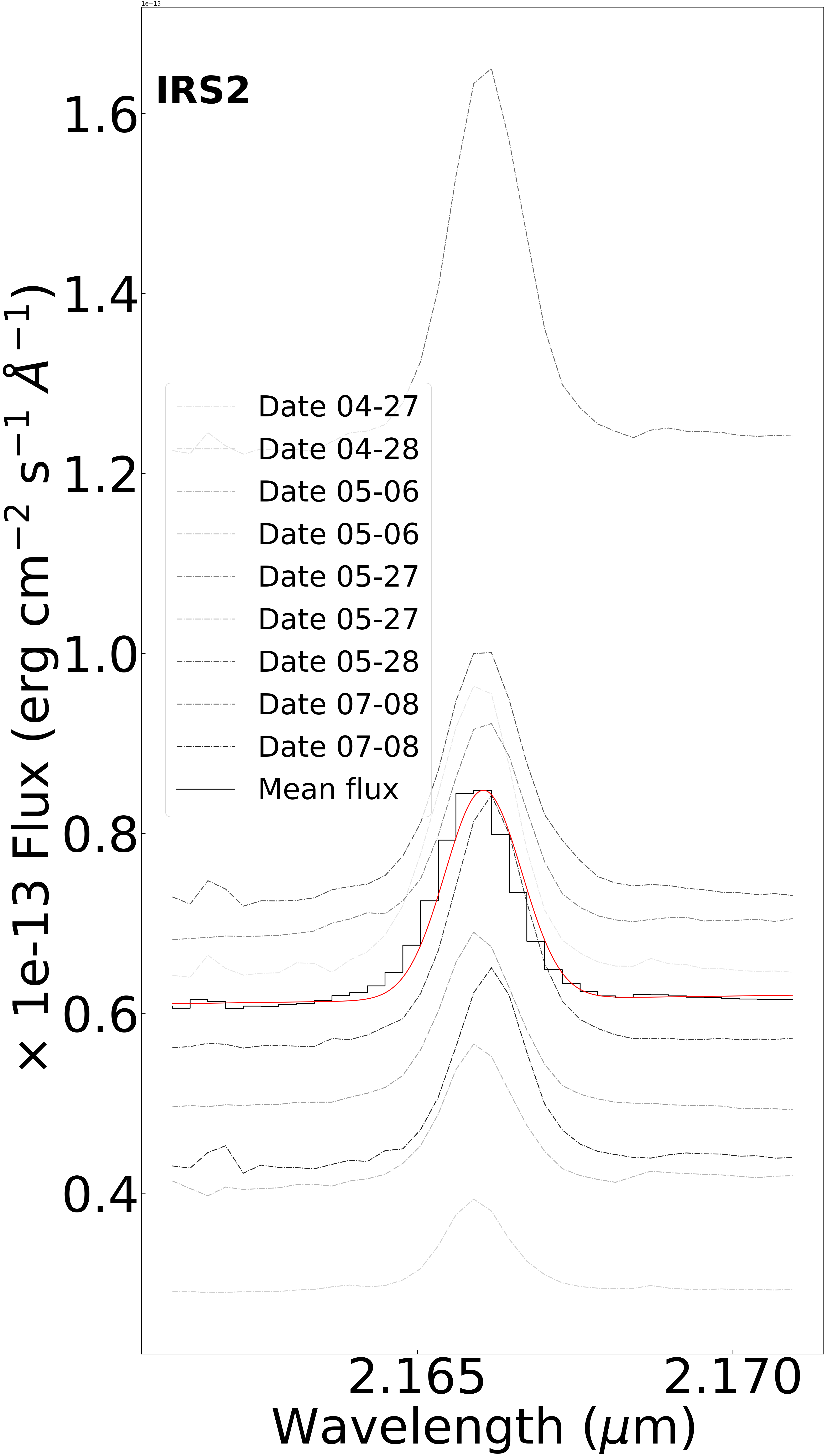}
    \end{subfigure}
    \begin{subfigure}[b]{0.22\textwidth}
        \includegraphics[width=\textwidth]{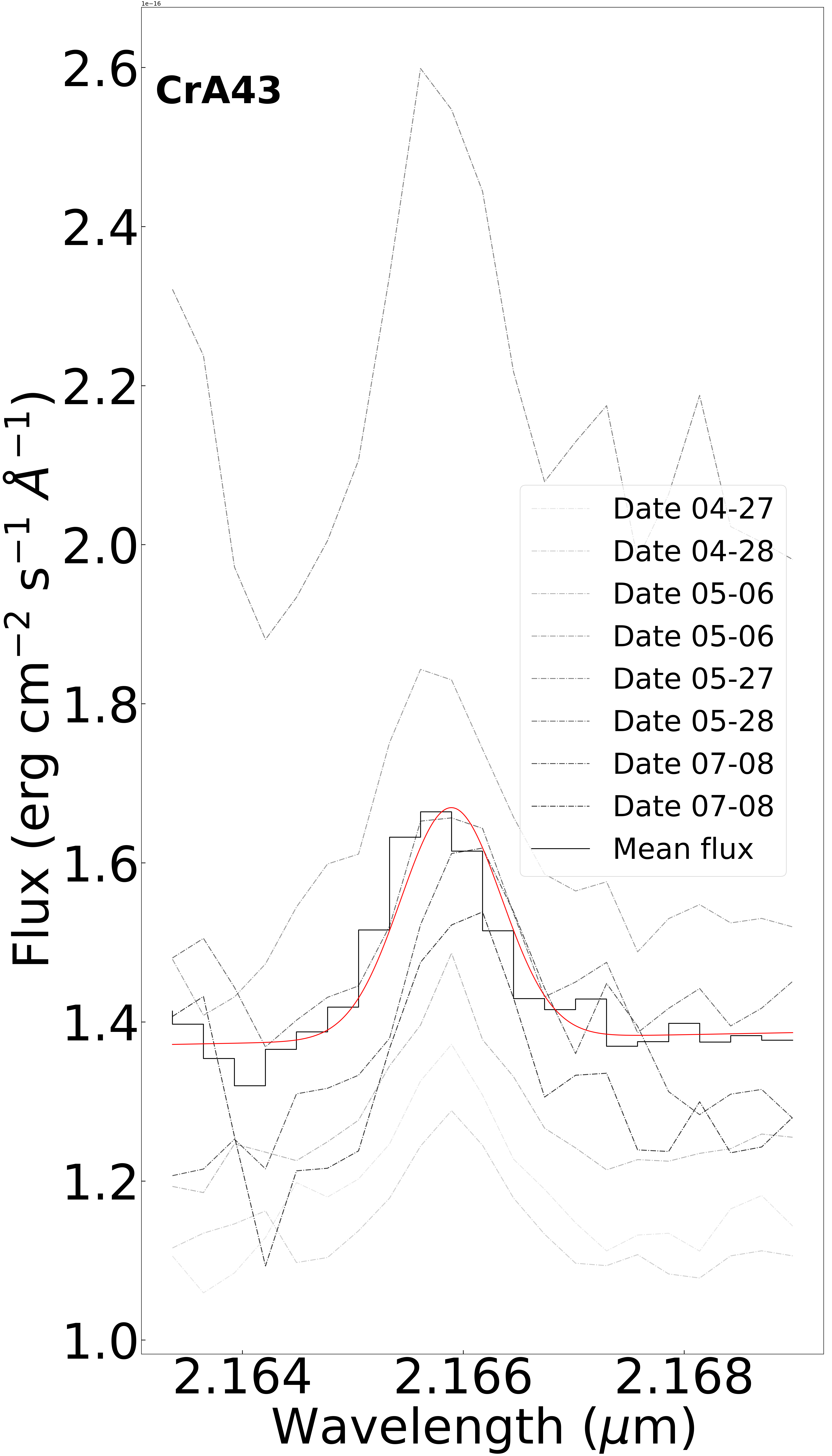}
    \end{subfigure}
    \begin{subfigure}[b]{0.22\textwidth}
        \includegraphics[width=\textwidth]{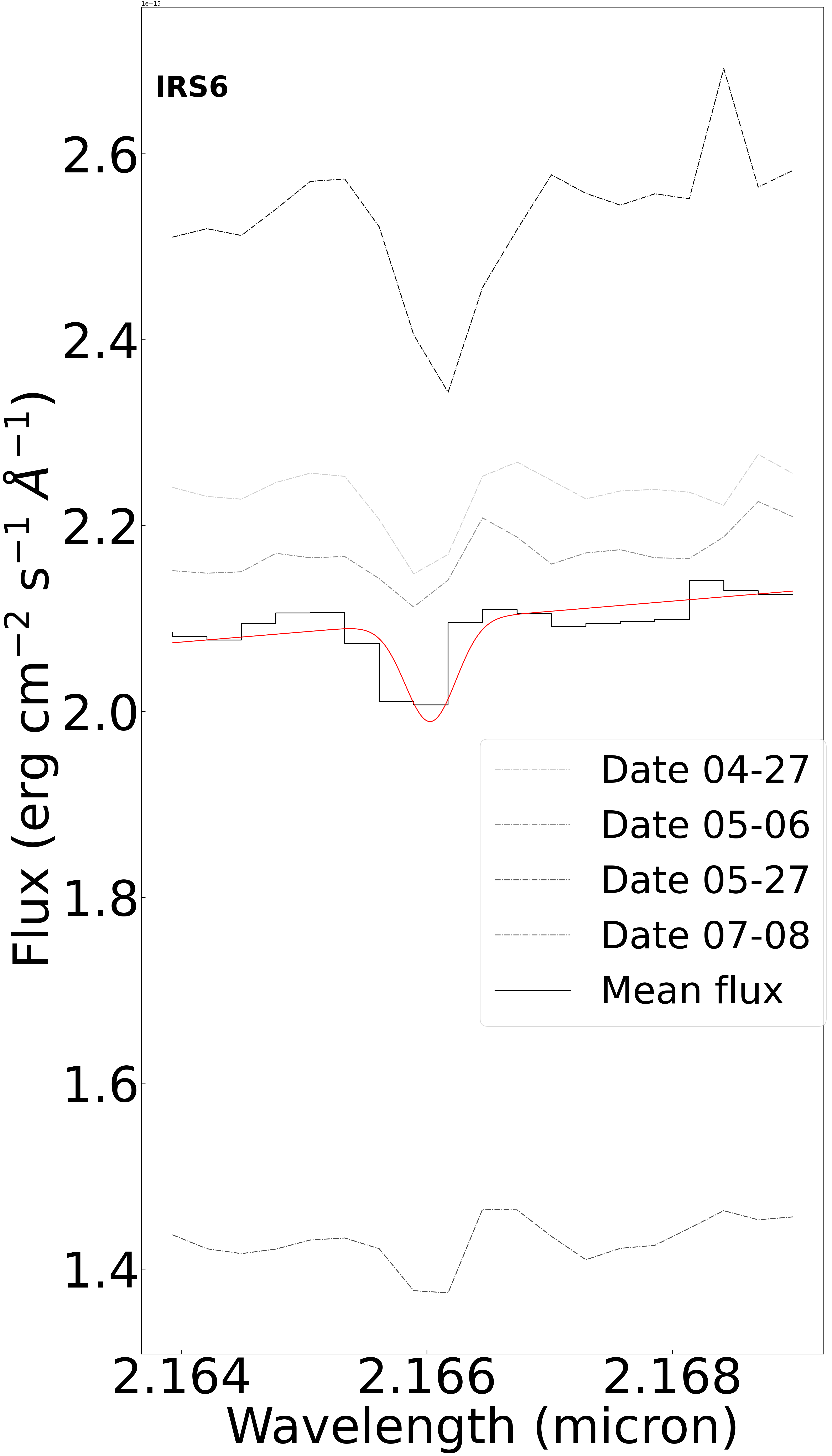}
    \end{subfigure}
    
    \vspace{0.2cm}  

    \begin{subfigure}[b]{0.22\textwidth}
        \includegraphics[width=\textwidth]{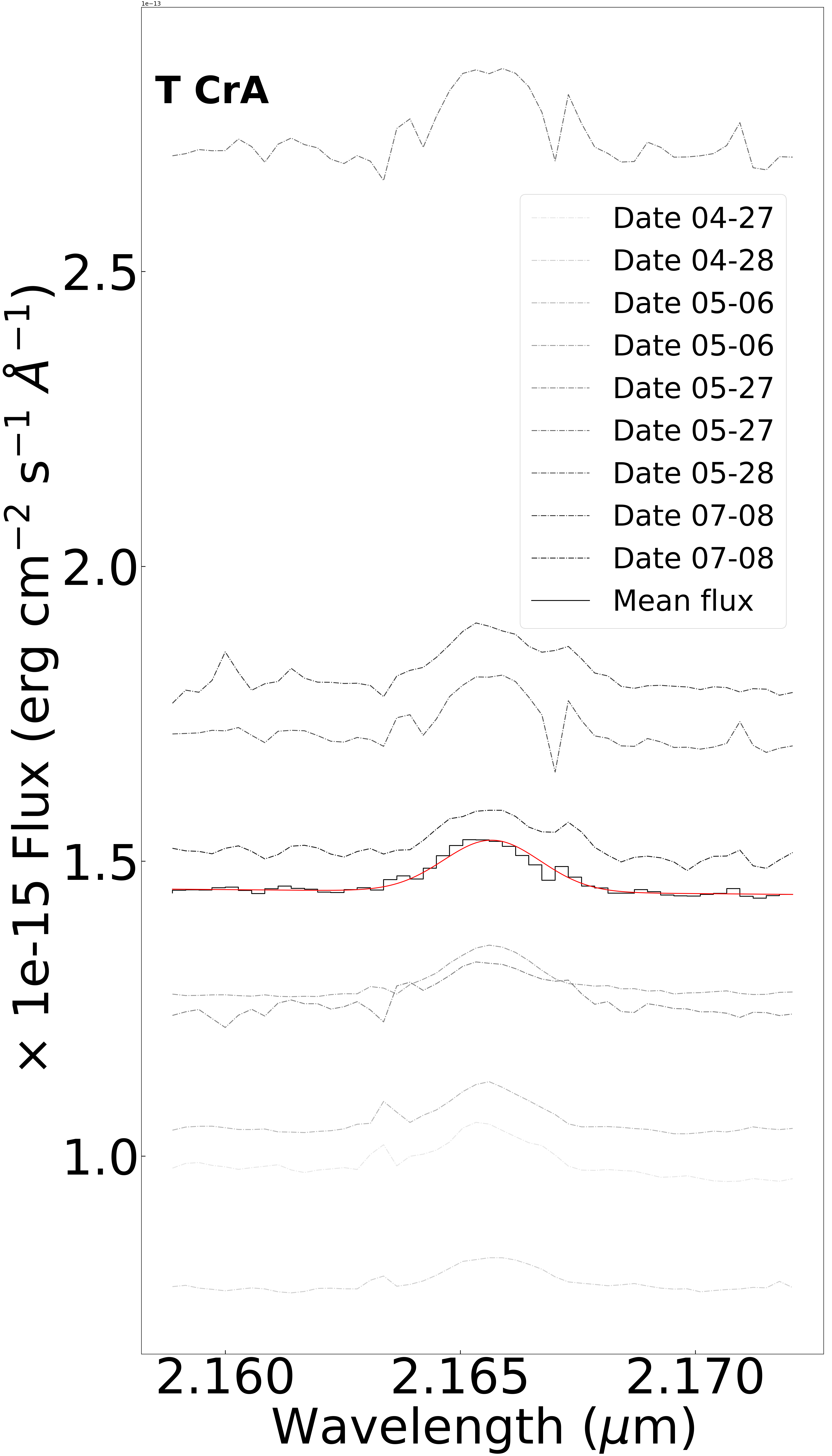}
    \end{subfigure}
    \begin{subfigure}[b]{0.22\textwidth}
        \includegraphics[width=\textwidth]{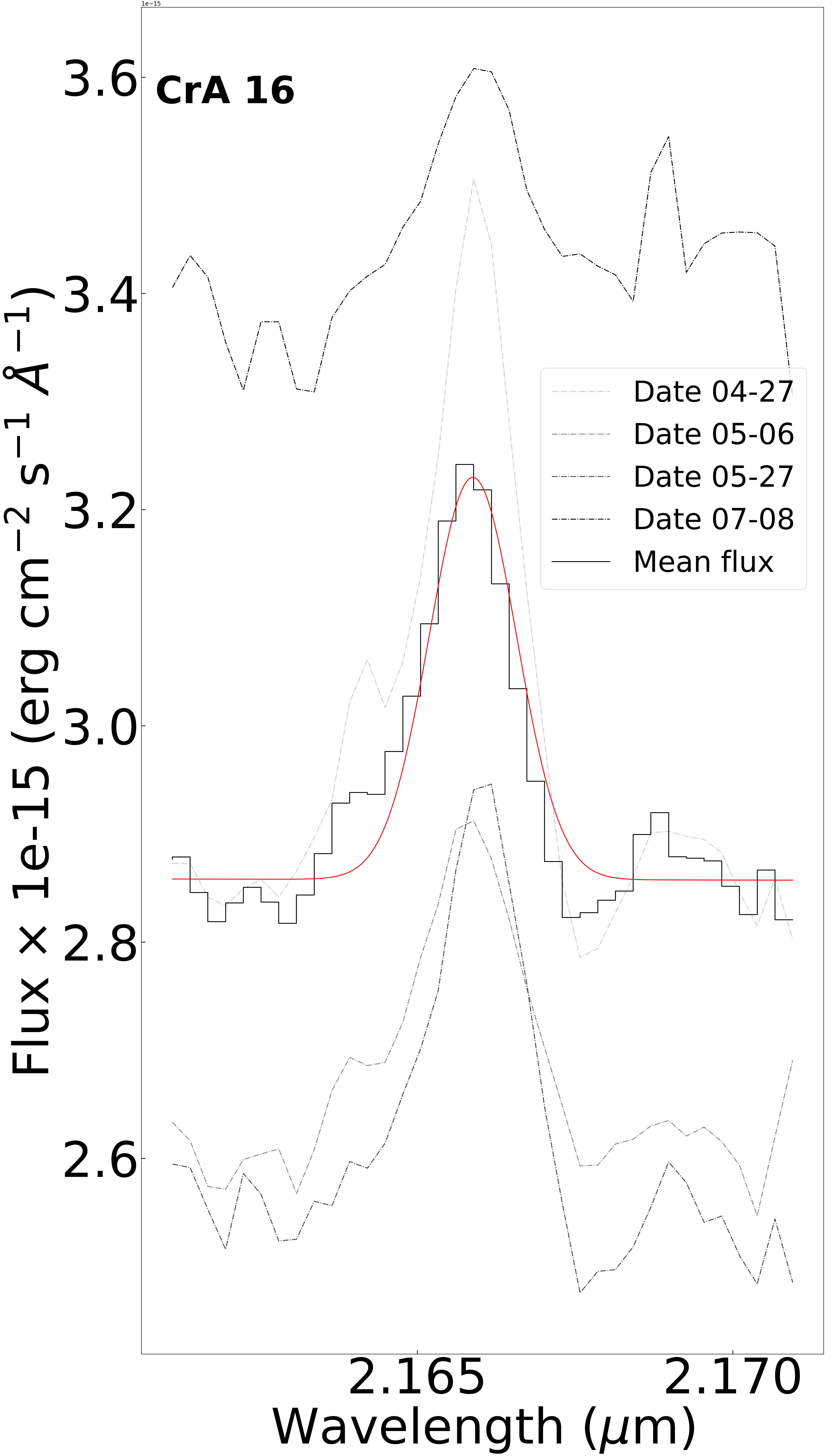}
    \end{subfigure}
    \begin{subfigure}[b]{0.22\textwidth}
        \includegraphics[width=\textwidth]{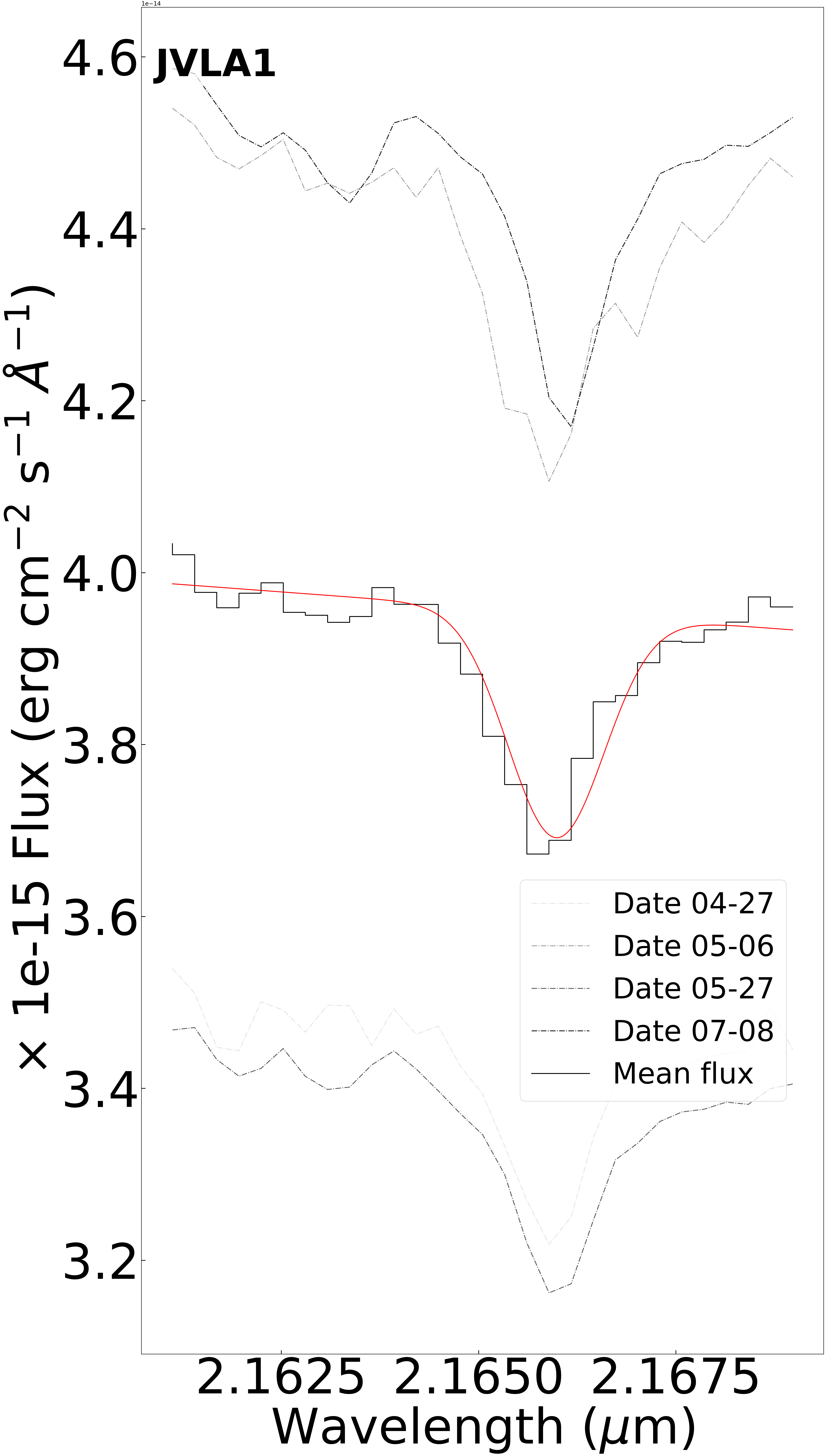}
    \end{subfigure}

    \caption{Br$\gamma$ line profiles of the seven YSOs with a detection listed in Table \ref{tab:source_data}. The broken grey lines show the spectra in individual epochs, whereas the step plot in black shows their average. The red line represents the Gaussian fit to the mean spectrum.}
    \label{fig:fig5}
\end{figure*}

\begin{figure*}
    \centering
    \includegraphics[width=0.3\textwidth]{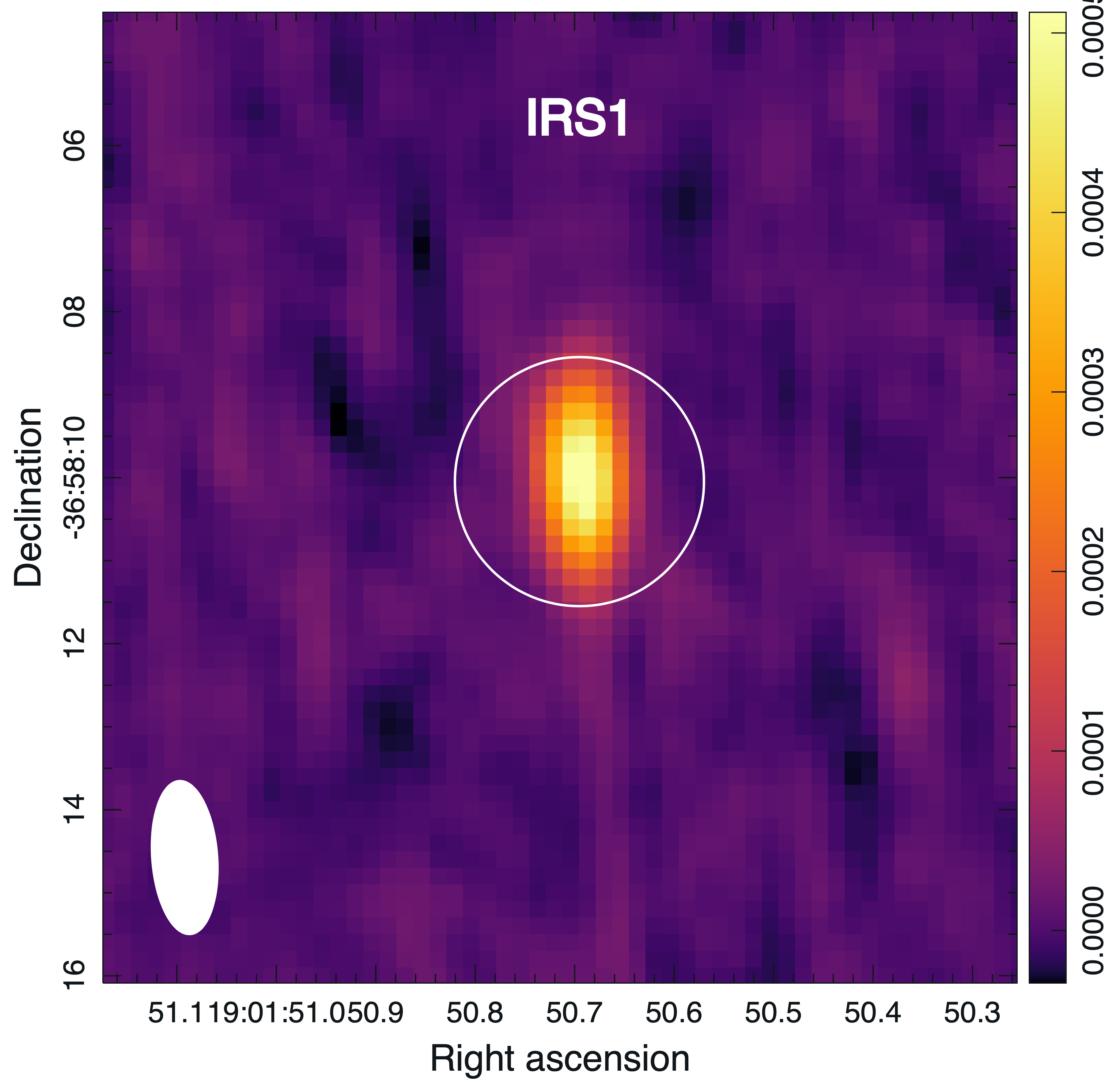}%
    \includegraphics[width=0.3\textwidth]{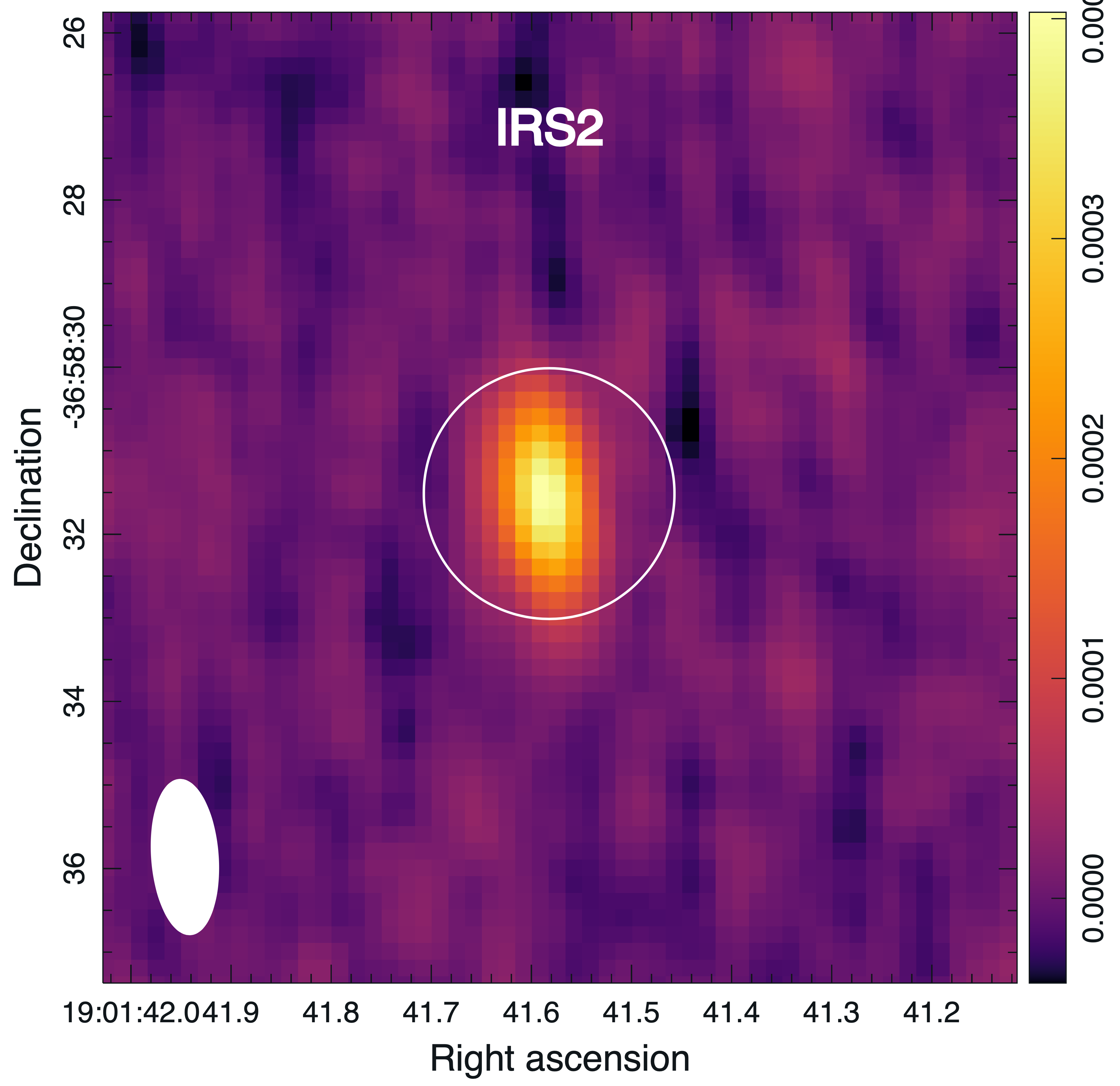}%
    \includegraphics[width=0.3\textwidth]{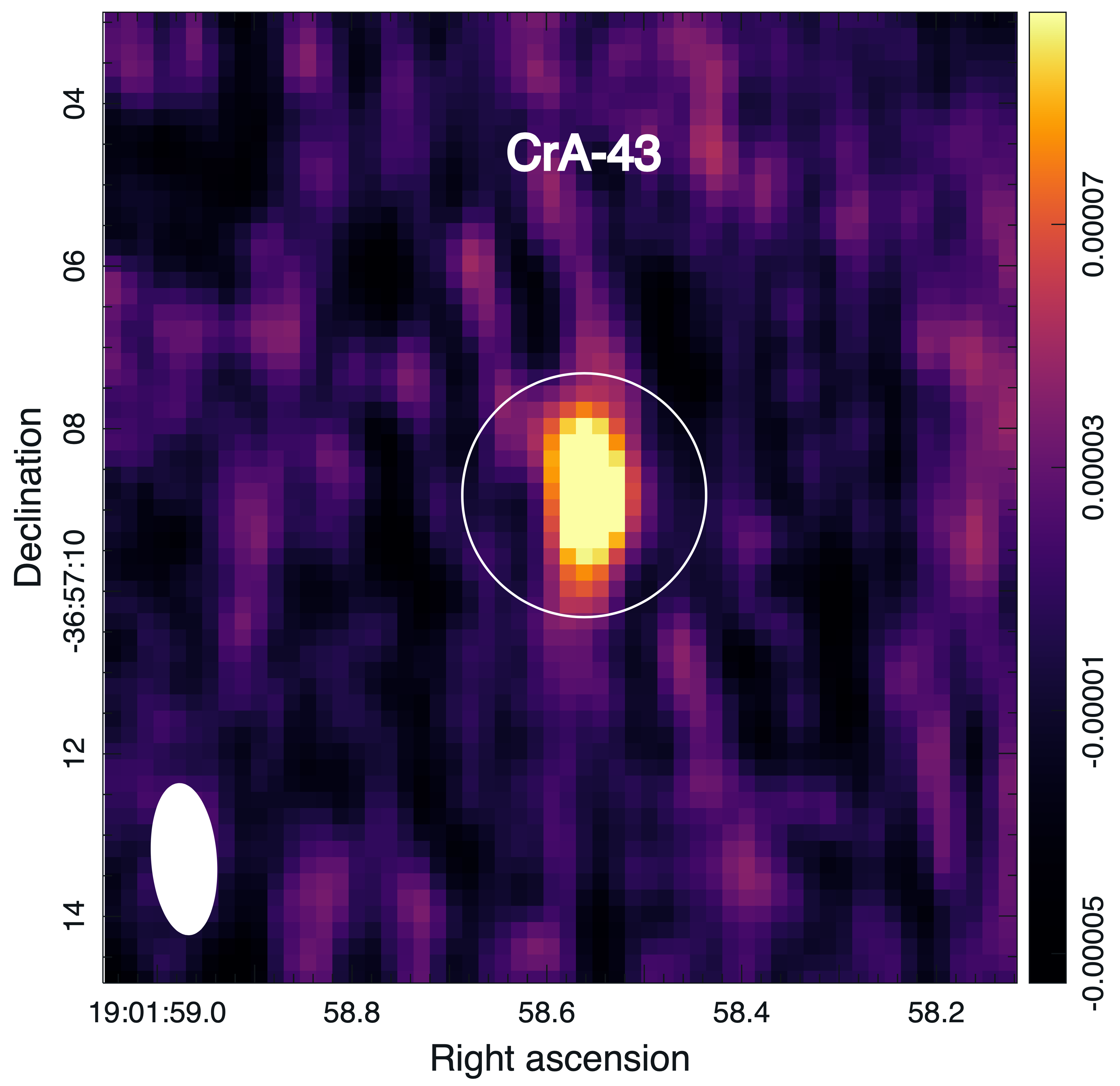}\par\vspace{0.2cm}
    \includegraphics[width=0.3\textwidth]{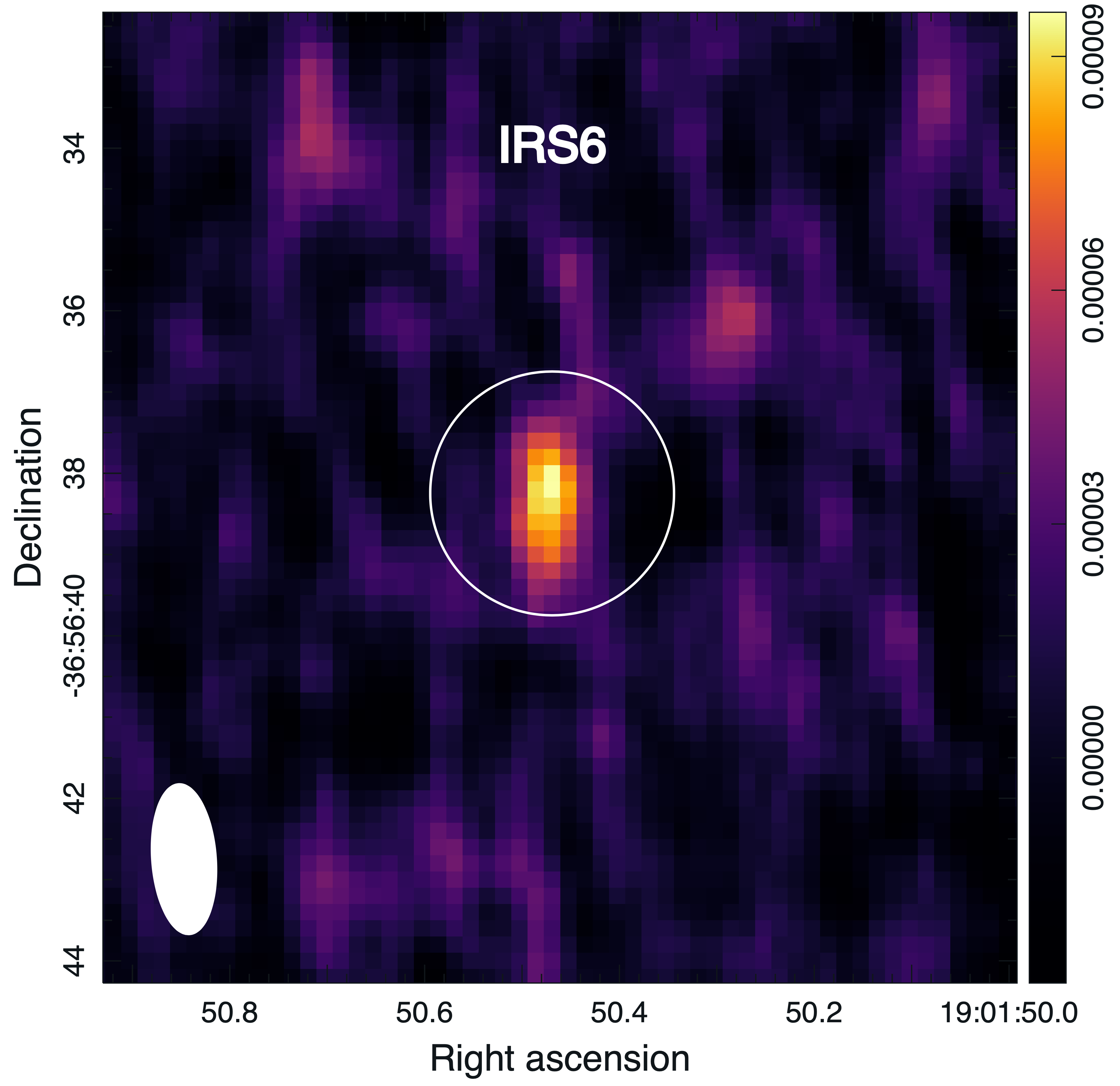}%
    \includegraphics[width=0.3\textwidth]{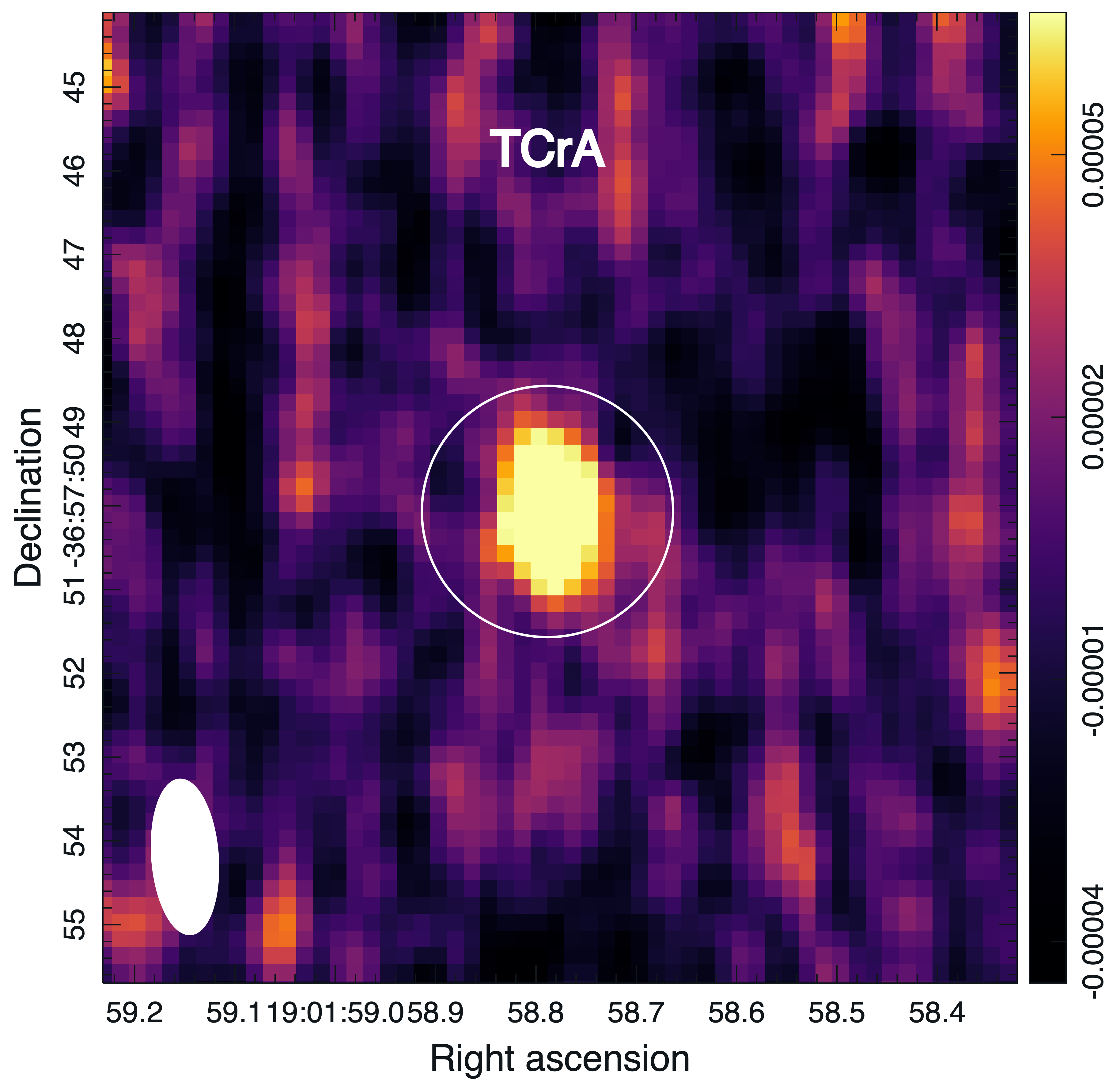}%
    \includegraphics[width=0.3\textwidth]{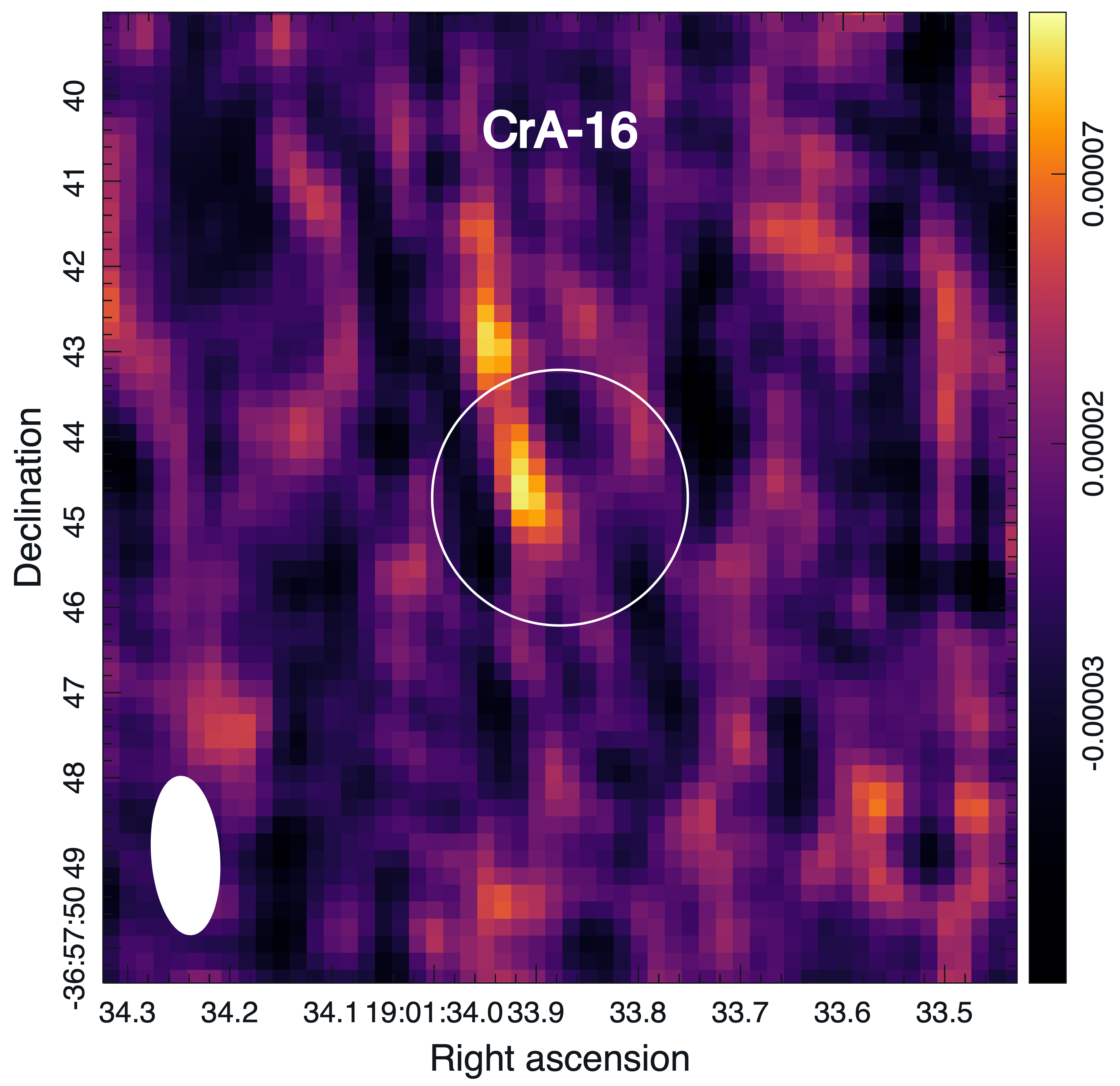}\par\vspace{0.2cm}
    \includegraphics[width=0.3\textwidth]{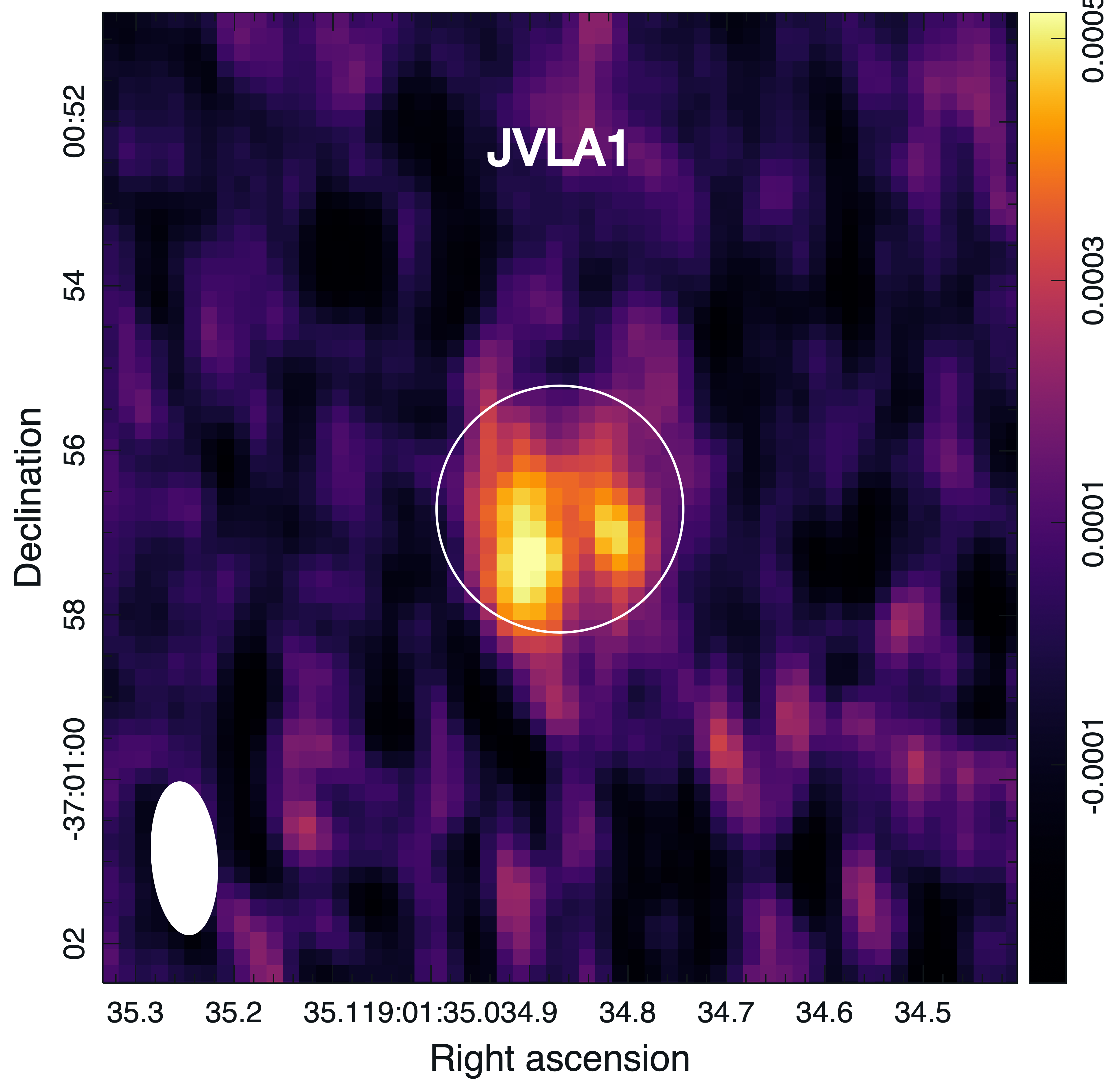}
    
    \caption{3.3 cm radio-continuum emission from the seven YSOs in the Coronet with Br$\gamma$ detection. The marker circles are centered at the coordinates listed in Table \ref{tab:source_data}. The color scale is normalized to the minimum and maximum intensities within each panel, using a square root color stretch.}
    \label{fig:radio_sources}
\end{figure*}

Figure \ref{fig:fig5} shows the VLT-KMOS detections of the Br$\gamma$ line listed in Table \ref{tab:source_data}, and Figure \ref{fig:radio_sources} shows the corresponding 3.3 cm radio continuum emission.

\newpage
\setcounter{table}{1}
\begin{table*}
\caption{Br$\gamma$ and 3.3 cm radio continuum fluxes for the YSOs listed in Table~\ref{tab:brg_radio_multiobs}. 
Columns from left to right are: source name, observation dates,
per-epoch Br$\gamma$ flux, per-epoch 3.3 cm flux, Br$\gamma$ flux from averaged spectrum, 
and 3.3 cm flux from the concatenated-data image. 
The fluxes marked with $^{\dagger}$ are upper limits.}
\label{tab:brg_radio_multiobs}
\centering
\begin{tabular}{llcccc}
\hline
Source & Date & $F_{\mathrm{Br}\gamma}$ & $F_{\mathrm{3.3cm}}$ & $\overline{F_{\mathrm{Br}\gamma}}$ & $\overline{F_{\mathrm{3.3cm}}}$ \\
 & 2014 & [erg s$^{-1}$ cm$^{-2}$] & [mJy] & [erg s$^{-1}$ cm$^{-2}$] & [mJy] \\
\hline
 & 03-17 & & 0.499 $\pm$ 0.019 & & \\
 & 04-22 & & 0.706 $\pm$ 0.019 & & \\
 & 04-26 & & 0.496 $\pm$ 0.011 & & \\
 & 05-03 & & 0.519 $\pm$ 0.019 & & \\
 & 05-06 & 2.16e-17 $\pm$ 3.19e-19 & & & \\
 & 05-06 & 2.31e-17 $\pm$ 7.33e-19 & & & \\ 
 IRS1 & 05-18 & & 0.508 $\pm$ 0.020 & 3.34e-17$\pm$6.48e-19 & 0.563 $\pm$ 0.009 \\
 & 05-27 & 5.69e-17 $\pm$ 1.64e-18 & & & \\
 & 07-06 & & 0.490 $\pm$ 0.042 & & \\
 & 07-08 & 3.22e-17 $\pm$ 9.91e-19 & & & \\
 & 07-11 & & 0.994 $\pm$ 0.039 & & \\        
 & 08-09 & & 0.530 $\pm$ 0.016 & & \\ 
 & 09-01 & & 0.519 $\pm$ 0.026 & & \\
\hline
 & 03-17 & & 0.371 $\pm$ 0.022 & & \\ 
 & 04-22 & & 0.303 $\pm$ 0.068 & & \\
 & 04-26 & & 0.377 $\pm$ 0.012 & & \\
 & 04-27 & 4.99e-17 $\pm$ 1.21e-18 & & & \\
 & 05-03 & & 0.418 $\pm$ 0.022 & & \\       
 IRS2 & 05-06 & 2.86e-17 $\pm$ 6.38e-19 & & 3.55e-17 $\pm$ 6.81e-19  & 0.399 $\pm$ 0.013 \\
 & 05-18 & & 0.346 $\pm$ 0.022 & & \\
 & 05-27 & 3.47e-17 $\pm$ 8.32e-19 & & & \\
 & 07-06 & & 0.385 $\pm$ 0.031 & & \\
 & 07-08 & 3.92e-17 $\pm$ 7.45e-19 & & & \\       
 & 07-11 & & 0.681 $\pm$ 0.104 & & \\
 & 08-09 & & 0.490 $\pm$ 0.040 & & \\ 
 & 09-01 & & 0.590 $\pm$ 0.021 & & \\
\hline
CrA16 & 04-27 & 1.02e-18 $\pm$ 9.52e-20 & & 6.45e-19 $\pm$ 5.40e-20 & 0.094 $\pm$ 0.021\\
 & 05-06 & 5.84e-19 $\pm$ 5.56e-20 & & & \\
 & 05-27 & 6.46e-19 $\pm$ 5.58e-20 & & & \\   
 & 07-08 & 3.49e-19 $\pm$ 7.81e-20 & & & \\
\hline
CrA26 & & & & 1.03e-20$^{\dagger}$ & 0.061$^{\dagger}$  \\ 
Peterson 1 & & & & 2.45e-20$^{\dagger}$ & 0.087$^{\dagger}$  \\
Peterson 6 & & & & 1.03e-21$^{\dagger}$ & 0.078$^{\dagger}$  \\
Peterson 7 & & & & 6.29e-21$^{\dagger}$ & 0.054$^{\dagger}$  \\
CrA9 & & & & 7.37e-21$^{\dagger}$ & 0.165$^{\dagger}$  \\
IRS10 & & &  & 7.68e-21$^{\dagger}$ & 0.084$^{\dagger}$  \\
Haas17 & & & & 4.53e-21$^{\dagger}$ & 0.072$^{\dagger}$  \\
\hline
\end{tabular}
\end{table*}





\bsp	
\label{lastpage}
\end{document}